\documentclass{aastex}
\usepackage{emulateapj5}

\newcommand{\kms}{km~s$^{-1}$}

\usepackage{psfig}
\usepackage{amsmath}
\usepackage{amssymb}
\usepackage{graphicx}

\received{} 
\accepted{} 
\journalid{}{} 
\articleid{}{} 
\shortauthors{D.~Kasen {\it et al.}}
\shorttitle{Spectropolarimetry of SN~2001el}

\begin{document}
\bibliographystyle{apj}

\title{Analysis of the Flux and Polarization Spectra of the\\
Type Ia Supernova SN~2001el:\\ 
Exploring the Geometry of the High-velocity Ejecta}

\author{Daniel~Kasen\altaffilmark{1}, Peter~Nugent\altaffilmark{1},
Lifan~Wang\altaffilmark{1}, D.A.~Howell\altaffilmark{1},
J.~Craig~Wheeler\altaffilmark{2}, Peter~H{\"o}flich\altaffilmark{2},
Dietrich~Baade\altaffilmark{3}, E.~Baron\altaffilmark{4},
P.H.~Hauschildt\altaffilmark{5}}   

\email{dnkasen@panisse.lbl.gov}

\altaffiltext{1}{Lawrence Berkeley National Laboratory, Berkeley, CA 94720}
\altaffiltext{2}{Department of Astronomy, University of Texas at
Austin, Austin, TX 78712}  
\altaffiltext{3}{European Southern Observatory,
Karl-Schwarzschild-Strasse 2, D-85748 Garching, Germany}
\altaffiltext{4}{Department of Physics and Astronomy, University of
Oklahoma, Norman, OK 73019}
\altaffiltext{5}{Department of Physics and Astronomy and Center for
Simulational Physics, University of Georgia, Athens, GA 30602} 

\begin{abstract}
SN~2001el is the first normal Type Ia supernova to show a strong,
intrinsic polarization signal. In addition, during the epochs prior to
maximum light, the CaII IR triplet absorption is seen distinctly and
separately at both normal photospheric velocities and at very high
velocities.  The high-velocity triplet absorption is highly
polarized, with a different polarization angle than the rest of the
spectrum.  The unique observation allows us to construct a relatively
detailed picture of the layered geometrical structure of the supernova
ejecta: in our interpretation, the ejecta layers near the photosphere
($v \approx 10,000$ \kms) obey a near axial symmetry, while a
detached, high-velocity structure ($v \approx 18,000-25,000$ \kms)
with high CaII line opacity deviates from the photospheric
axisymmetry.  By partially obscuring the underlying photosphere, the
high-velocity structure causes a more incomplete cancellation of the
polarization of the photospheric light, and so gives rise to the
polarization peak and rotated polarization angle of the high-velocity
IR triplet feature.  In an effort to constrain the ejecta geometry, we
develop a technique for calculating 3-D synthetic polarization spectra
and use it to generate polarization profiles for several parameterized
configurations.  In particular, we examine the case where the inner
ejecta layers are ellipsoidal and the outer, high-velocity structure
is one of four possibilities: a spherical shell, an ellipsoidal shell,
a clumped shell, or a toroid.  The synthetic spectra rule out the
spherical shell model, disfavor a toroid, and find a best fit with the
clumped shell.  We show further that different geometries can be more
clearly discriminated if observations are obtained from several
different lines of sight.  Thus, assuming the high velocity structure
observed for SN~2001el is a consistent feature of at least a known
subset of type Ia supernovae, future observations and analyses such as
these may allow one to put strong constraints on the ejecta geometry
and hence on supernova progenitors and explosion mechanisms.
\end{abstract}

\section{Introduction}
\subsection{Spectropolarimetry of Supernova}
The geometrical structure of supernova ejecta, as determined
empirically from observations, can give important clues as to the
nature of the supernova progenitor system and explosion physics.
Spectropolarimetry is a crucial tool in constraining the shape of
unresolved supernovae. The scattering atmospheres found in supernovae
can linearly polarize light. For an unresolved, spherically symmetric
system the differently aligned polarization vectors around the disk
will cancel, resulting in zero net polarization. If the symmetry
around the line of sight is broken, however, a net polarization can
result due to incomplete cancellation of polarization vectors
\citep{shapsuth}.

The polarization observations of SN~2001el presented in
\citet{wang_pol_01el} (hereafter Paper~I) are the first observations
of a spectroscopically normal Type~Ia supernova (SN~Ia) which show a
significant intrinsic polarization signal.  Most previous observations of
SN~Ia showed no observable polarization, given the signal-to-noise of
the observations \citep{wang_pol_96}.  The only other indication of
a clear non-zero polarization in a SN~Ia was the subluminous and
spectroscopically peculiar SN~Ia 1999by, which showed an intrinsic
continuum polarization of about 0.7$\%$ \citep{Howell-99by}.  Chemical
inhomogeneities were also suggested to explain the rather noisy
polarization data of SN~1996x \citep{wang-96x}. In addition, strong
intrinsic polarization has been measured in all types of core collapse
supernovae \citep{wang_pol_96}.

A non-zero intrinsic polarization measurement indicates that a
supernova is aspherical, but using the spectropolarimetry to constrain
the supernova geometry usually requires theoretical modeling. The
detailed theoretical studies so far have been confined to axisymmetric
configurations. \citet{shapsuth} first estimated the continuum
polarization expected from an ellipsoidal, electron scattering
supernova atmosphere. \citet{Hoflich-91-pol} used a Monte Carlo code
to calculate the continuum polarization from several axisymmetric
configurations, including an off-center energy source embedded in a
spherical electron scattering envelope. Calculations of synthetic
supernova polarization spectra have also been performed, but usually
only for the ellipsoidal geometries (see however
\citet{chugai_87Apol}). In the past, such ellipsoidal models have done
a fair job in fitting gross characteristics of the available
spectropolametric observations, for example those of SNe~1987A
\citep{Jeffery-87A}, 1993J \citep{Hoflich-93J} and SN~1999by
\citep{Howell-99by}.

SN~2001el presents an exciting development in that no axially
symmetric geometry is able to account entirely for the
spectropolametric observations.  In particular, we suggest that the
supernova ejecta consists of nearly axially-symmetric inner layers ($v
\la 15,000$ \kms), surrounded by a detached, high-velocity structure
($v \approx 20,000-25,000$ \kms) with a different orientation.  The
analysis of the system therefore requires that we consider the
synthesis of polarization spectra for 3-D configurations.

In this paper we take an empirical approach, and use a parameterized
model to try to extract as much \emph{model independent} information
about the high velocity structure in SN~2001el as the observations
will permit. A unique 3-D reconstruction of the geometry is not
possible, as this constitutes a kind of ill-posed inverse problem.
However, by restricting our attention to various parameterized
systems, we can draw some rather general conclusions about the
viability of different geometries.  In particular, we examine the case
where the inner ejecta layers are ellipsoidal and the outer,
high-velocity structure is one of four possibilities: a spherical
shell, an ellipsoidal shell, a clumped shell, or a toroid.  We develop
a technique for calculating 3-D synthetic polarization spectra of the
high velocity material. The synthetic spectra rule out the spherical
shell model, disfavor a toroid, and find a best fit with the clumped
shell.

Geometrical information extracted empirically from spectropolarimetry
must eventually be compared to detailed multi-dimensional explosion
models.  As of yet, none of the computed explosion models appear
directly applicable to SN~2001el.  3-D deflagration models of a SN~Ia
in the early phases have been computed by \citet{khokhlov_3d} and
\citet{Reinecke_3d}.  These models show a quite inhomogeneous chemical
structure, with large plumes of burned material extending into
unburned material. So far the calculations only cover the early stages
of the explosion, before free expansion is reached.  It is possible
that at some point the deflagration transitions into a detonation wave
\citep{khok91a}.  The detonation may smooth out the inhomogeneities in
the chemical composition by burning away the unburnt material between
the plumes \citep{Hoeflich_IR, khokhlov_3d}.  It could also introduce
a global asymmetry if it occurs at an off-center point
\citep{Livne_99}.  Other possible sources of asymmetry include rapid
rotation of a white dwarf progenitor \citep{Mahaffy}, and the binary
nature of the progenitor system \citep{Marietta}.

\subsection{Supernova SN~2001el}
\citet{sn01el_iau} discovered SN~2001el in the galaxy NGC 1448. The
brightness of this nearby supernova ($m_{B} \approx 12$ at peak) made
it an ideal candidate for spectropolarimetry. Spectropolametric
observations were taken on Sept 25, Sept 30, Oct 9 and Nov 9 of
2001. Details on the observations and the data reduction of the
spectra analyzed in this paper can be found in Paper I.

\begin{figure}
\begin{center}
\leavevmode
\psfig{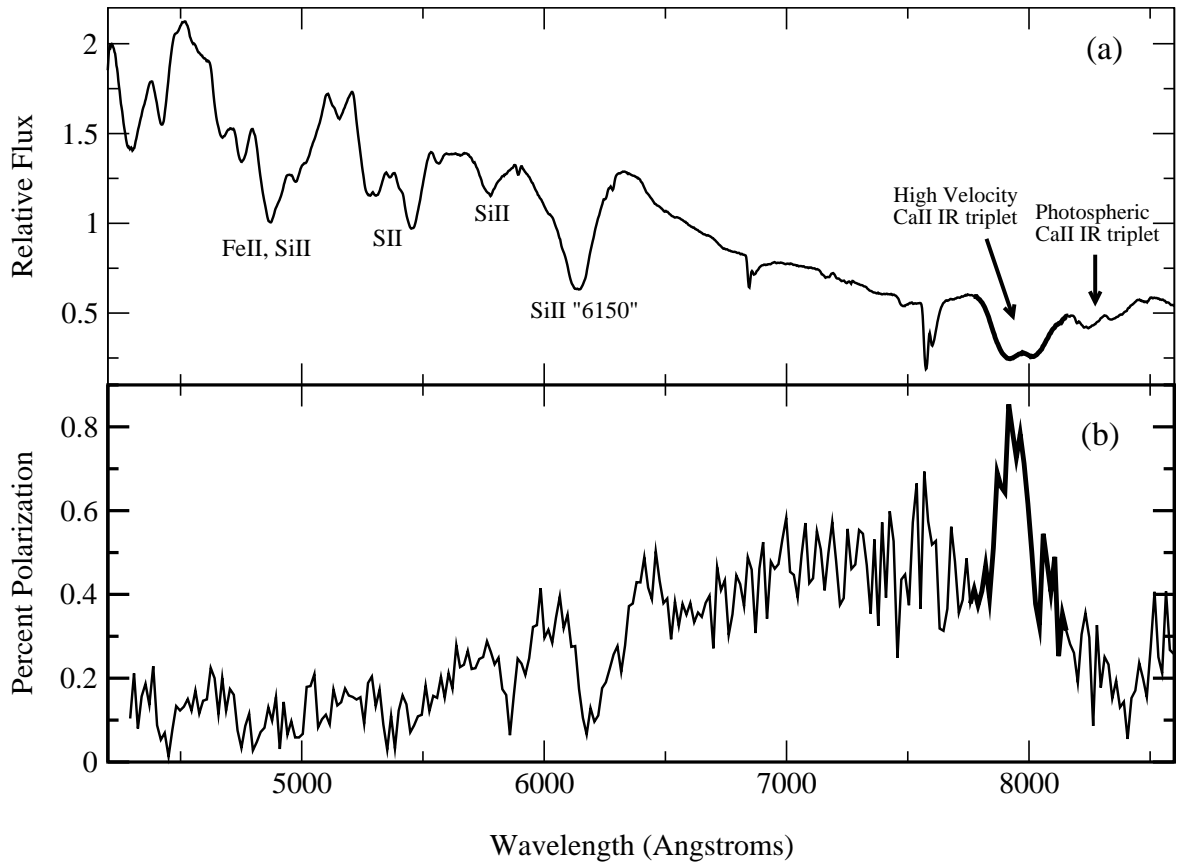}
\caption{Flux and polarization spectrum of SN~2001el on Sept 25. The
HVM feature is shown in bold lines.  The polarization spectrum has
been ISP subtracted using the ISP shown as the square in
Figure~\ref{qu_plot}.
\label{spec_plot}}
\end{center}
\end{figure}

In Figure~\ref{spec_plot}a we show the flux spectrum of SN~2001el for
the first epoch (we have removed the redshift due to the peculiar
velocity of the host galaxy). The flux spectrum of SN~2001el resembles
the normal SN~Ia SN~1994d at about 7 days before maximum light, with
the expected P-Cygni features due to \ion{Si}{2}, \ion{S}{2},
\ion{Ca}{2} and \ion{Fe}{2} (see e.g. \citet{bfn93}). The blueshifts
of the minima of these features can be used to estimate the
photospheric velocities of SN~2001el, which for all features are found
to be $v_{ph} \approx 10,000$ \kms.  The only truly unusual feature of
the flux spectrum is a strong absorption near 8000~\AA, which is 
discussed in detail below.

We concentrate our analysis on the earliest spectrum (Sept.~25), of
SN~2001el.  A full description of the flux and polarization spectra at all
epochs is given in Paper I.

\subsection{High Velocity Material in SNe~Ia}
The most interesting feature of SN~2001el is the strong absorption
feature near 8000~\AA. The absorption has a ``double-dipped'' profile,
consisting of two partially blended minima separated by about
150~\AA. It seems to be a pure absorption feature with no obvious
emission component to the red. The feature is still strong on Sept 30,
but has weakened considerably by Oct 9. By the Nov 9 observations, the
8000~\AA\ feature has virtually disappeared (see Paper~I).

\citet{Hatano94D99} identified a much weaker 8000~\AA\ feature in
SN~1994D as a highly blueshifted \ion{Ca}{2} IR triplet. The
double-dipped profile now visible in the Sept 25 SN~2001el spectrum
supports this conclusion.  The red-most line of the triplet
($\lambda$8662) produces the red-side minimum while the two other
triplet lines ($\lambda$8542 \& $\lambda$8498) blend to produce the
blue-side minimum.  The synthetic spectra to be presented in
\S\ref{geohvm} confirm that the IR triplet can reproduce the shape of
the double minimum. Unfortunately, the early spectra do not extend far
enough to the blue to observe a corresponding high velocity component
to the \ion{Ca}{2} H\&K lines.  We have investigated all other
potential lines that might have caused the 8000~\AA\ feature, but none
were able to reproduce the feature without producing another
unobserved line signature somewhere else in the spectrum.

Adopting the IR triplet identification for the 8000~\AA\ feature, the
implied calcium line of sight velocities span the range $18,000 -
25,000$ \kms. This should be contrasted with the photospheric velocity
of 10,000 \kms~as measured from the normal SN~Ia features in
SN~2001el. We therefore make the distinction between the
\emph{photospheric material}, which gives rise to a seemingly normal
SN~Ia spectrum (hereafter, the ``photospheric spectrum''), and the
\emph{high velocity material} (HVM), which produces the unusual
8000~\AA\ IR triplet feature. In the flux spectrum, there is a clear
separation between the photospheric triplet absorption at 8300~\AA\
and the HVM feature at 8000~\AA. In the polarization spectrum, the
angle and degree of polarization of the 8000~\AA\ feature each differ
from the photospheric spectrum. Both of these imply a rather sudden
change of the atmospheric conditions in the HVM.

A high velocity CaII IR triplet feature has been observed in other
SNe~Ia, albeit rarely and never as strong. The pre-max spectra of
SN~1994D \citep{patat94d,meik94d}, show a similar, but much weaker
absorption.  The \ion{Si}{2} and \ion{Fe}{2} lines of these spectra
also suggest some material is moving faster than 25,000 \kms~
\citep{Hatano94D99} The earliest spectrum of SN~1990N at day -14
\citep{Leib_90n} has a deep, rounded 8000~\AA\ feature, and the
spectrum also showed evidence of high velocity silicon or carbon
\citep{fish90n97}.  The 8000~\AA\ feature has also been observed in
the maximum light spectrum of SN~2000cx \citep{li_00cx}. In this case,
however, the line widths are narrower and the two minima are almost
completely resolved. 

In SN~2001el, the only clear-cut evidence for high velocity material
seems to be the 8000~\AA\ feature. There is no strong \ion{Si}{2} 6150
absorption at $v > 20,000$, although a weak absorption cannot be ruled
out because at this wavelength (5880~\AA) it would blend completely
with the neighboring \ion{Si}{2} $\lambda\lambda5958,5979$ feature.
There is also no clear indication of high velocity \ion{Fe}{2} or
\ion{S}{2}.  The blue edge of the \ion{Ca}{2} H\&K feature on Oct.~9
-- the first available spectrum to go far enough to the blue -- is at
27,000 \kms. The likelihood of this being HVM is suspect because of
the strong possibility of line blending. Since the 8000~\AA\ feature
is the only unambiguous detection of a high velocity material in
SN~2001el, we hereafter refer to it as \emph{the} HVM feature.

Our analysis will focus almost entirely on the 8000~\AA\ HVM
feature. In \S\ref{specpol} we give an introduction to polarization in
supernova atmospheres; \S\ref{twocomp} describes a parameterized model
that allows us to generate synthetic polarization spectra, and in
\S\ref{geohvm} we use the model to explore various geometries for
SN~2001el. In \S\ref{los} we consider the signature of each geometry
when viewed from alternative lines of sight.  The implication of these
constraints on the progenitors and explosion mechanisms of SNe~Ia's is
discussed briefly in the conclusion.

\section{Supernova Spectropolarimetry}
\label{specpol}
\subsection{Polarization Basics}
The polarization state of light describes an anisotropy in the
time-averaged vibration of the electric field vector. A beam of radiation
where the electric field vector vibrates in one specific plane is
completely (or fully) linearly polarized. A beam of radiation where the
electric field vector vibrates with no preferred direction is 
unpolarized. Imagine holding a polarization filter in front of a
completely linearly polarized light beam of intensity $I_0$. The
filter only transmits the component of electric field parallel to the filter
axis. Thus as the filter is rotated, the transmitted intensity, which
is proportional to the square of the electric field, varies as
$I(\theta) = I_0 cos^2\theta$.

The light measured from astrophysical objects is the superposition of
many individual waves of varying polarization. Imagine a light
beam consisting of the super-position of two completely linearly
polarized beams of intensity $I_0$, and $I_{90}$, whose electric field
vectors are oriented 90$^\circ$ to each other. If the beams add
incoherently, the transmitted intensity is the sum of each
separate beam intensity:
\begin{equation}
\begin{split}
I(\theta) =& I_0\cos^2\theta + I_{90}\cos^2(\theta + 90^\circ) 
\\ &= I_0\cos^2\theta + I_{90} \sin^2\theta
\end{split}
\end{equation}
If the beams are of equal intensity, $I_0 = I_{90}$, then the
transmitted intensity shows no directional dependence upon $\theta$ --
i.e. the light is unpolarized. In this sense, we say that the
polarization of a light beam is ``canceled'' by an equal intensity
beam of orthogonal -- or ``opposite'' -- polarization. If $I_0 \neq
I_{90}$ the cancellation is incomplete, and the beam is said to be
\emph{partially polarized}.  The degree of polarization $P$ is defined
as the maximum percentage change of the intensity; in this case:
\begin{equation}
 P  = \frac{(I_0 - I_{90})}{I_0 + I_{90}} 
\end{equation}
The polarization position angle (labeled $\chi$) is defined as the
angle at which the transmitted intensity is maximum.

It is tempting to think of the polarization as a (two dimensional)
vector, since it has both a magnitude and a direction. Actually the
polarization is a percent difference in intensity, and intensity is
the \emph{square} of a vector (the electric field). The polarization
is actually a quasi-vector, i.e. polarization directions 180$^\circ$
(not 360$^\circ$) apart are considered identical.  The additive
properties of the polarization thus differ slightly from the vector
case, as evidenced by the fact that the polarization is canceled by
another equal beam oriented $90^\circ$ to it, rather than one at
$180^\circ$ as in vector addition.

In this case, a useful convention for describing polarization is
through the \emph{Stokes Parameters}, $I, Q$ and $U$, which measure
the difference of intensities oriented 90$^\circ$ to each other.  A
Stokes ``Vector'' can be defined and illustrated pictorially as:
\begin{equation}
\bf{I} = \begin{pmatrix}I\\Q\\U\end{pmatrix}
= \begin{pmatrix}I_{0^\circ} + I_{90^\circ} \\
		I_{0^\circ}  - I_{90^\circ  }\\ 
		I_{45^\circ} - I_{-45^\circ}
\end{pmatrix}
 = \begin{pmatrix}\updownarrow + \leftrightarrow\\
\updownarrow - \leftrightarrow\\
\searrow - \nearrow \\
\end{pmatrix}
\label{stokes_eq}
\end{equation}
where $I_{90^\circ}$, for instance, designates the intensity measured
with the polarizing filter oriented 90$^\circ$ to a specified
direction called the polarization reference direction. To determine
the superposition of two polarized beams, one simply adds their Stokes
vectors. A fourth Stokes parameter $V$ measures the excess of circular
polarization in the beam.  Non-zero circular polarization has not been
measured in supernova, and no circular polarization observations were
taken for SN~2001el; therefore we will not discuss Stokes $V$ in this
paper.  For scattering atmospheres without magnetic fields, the
radiative transfer equation for circular polarization separates from
the linear polarization equations, allowing us to ignore $V$ in our
calculations \citep{chandra60}.

We further define the fractional polarizations: $q = Q/I$ and $u=U/I$.
The degree of polarization, $P$, and the position angle $\chi$ can
then be written in terms of the Stokes Parameters:
\begin{equation}
\begin{split}
P = \frac{\sqrt{Q^2 + U^2}}{I}  = \sqrt{q^2 + u^2}\\
\chi = \frac{1}{2}\tan^{-1} (U/Q) = \frac{1}{2}\tan^{-1} (u/q)
\end{split}
\label{Pol_Stokes}
\end{equation}

A single plot that captures both the change of polarization degree and
position angle over a spectrum is the q-u plot of
Figure~\ref{qu_plot}. Each point in this figure is a wavelength
element of the spectrum, and for each point we can read off $P$ and
$\chi$ at that wavelength much as we would read a polar plot.
According to Equation~\ref{Pol_Stokes}, the degree of polarization $P$
is given by the distance of the point from the origin, while the
position angle $\chi$ is \emph{half} that of the plot's polar angle.
In this sense $q$ and $u$ can be thought of as the two components of a
two dimensional polarization quasi-vector.

\subsection{Polarization in Supernova Atmospheres}
The major opacities in a supernova atmosphere are due to electron
scattering and bound-bound line transitions.  The continuum
polarization of supernova spectra is attributed to electron
scattering.  The line opacity can create features (either peaks or
troughs) in the polarization spectra.

To understand the polarizing effect of an electron scattering, note
that an electron scatters a fully polarized beam of radiation
according to dipole $\sin^2\psi$ angular distribution, where $\psi$ is
the angle measured from the incident polarization direction.  Now
unpolarized light can be represented by a super-position of two equal
intensity, fully-polarized orthogonal beams. Upon electron scattering,
the two differently oriented beams get redistributed according to
differently oriented dipole patterns; thus in certain directions they
are no longer equal and do not cancel.  The scattered light is
therefore polarized with the percent polarization depending upon the
scattering angle $\Theta$ between incident and scattered rays:
\begin{equation}
P = \frac{1 - \cos^2\Theta}{1 + \cos^2\Theta}
\end{equation}
Light scattered at 90$^\circ$ is fully polarized, while that which is
forward scattered at 180$^\circ$ remains unpolarized. The direction of
the polarization is perpendicular to the scattering plane defined by
the incoming and outgoing photon directions.

Deep enough within the supernova atmosphere, the light becomes
unpolarized for two reasons: (1) Below a certain radius, known as the
\emph{thermalization depth}, the absorptive opacity dominates the
scattering opacity and photons are destroyed into the thermal
pool. The energy is subsequently re-emitted as blackbody radiation
which, being the result of random collision processes, is necessarily
unpolarized.  (2) Deep within the atmosphere, the radiation field
becomes isotropic.  Because the radiation incident on a scatterer is
then equal in all directions, the net polarization of scattered light will
cancel.

The polarization of the radiation occurs above the inner
unpolarized depth, where the election scattering opacity dominates and
the radiation field becomes anisotropic due to the escape of photons
out of the supernova surface.  We call this region the
electron-scattering zone.  The surface above the electron scattering
zone at which point photons have a high probability of escaping the
atmosphere, is the supernova \emph{photosphere}. Formation of the
well-know P-Cygni line profiles in supernovae is due to line opacity
from material primarily above the photosphere. This region is called
the line-forming region.

\begin{figure}
\begin{center}
\leavevmode
\psfig{file=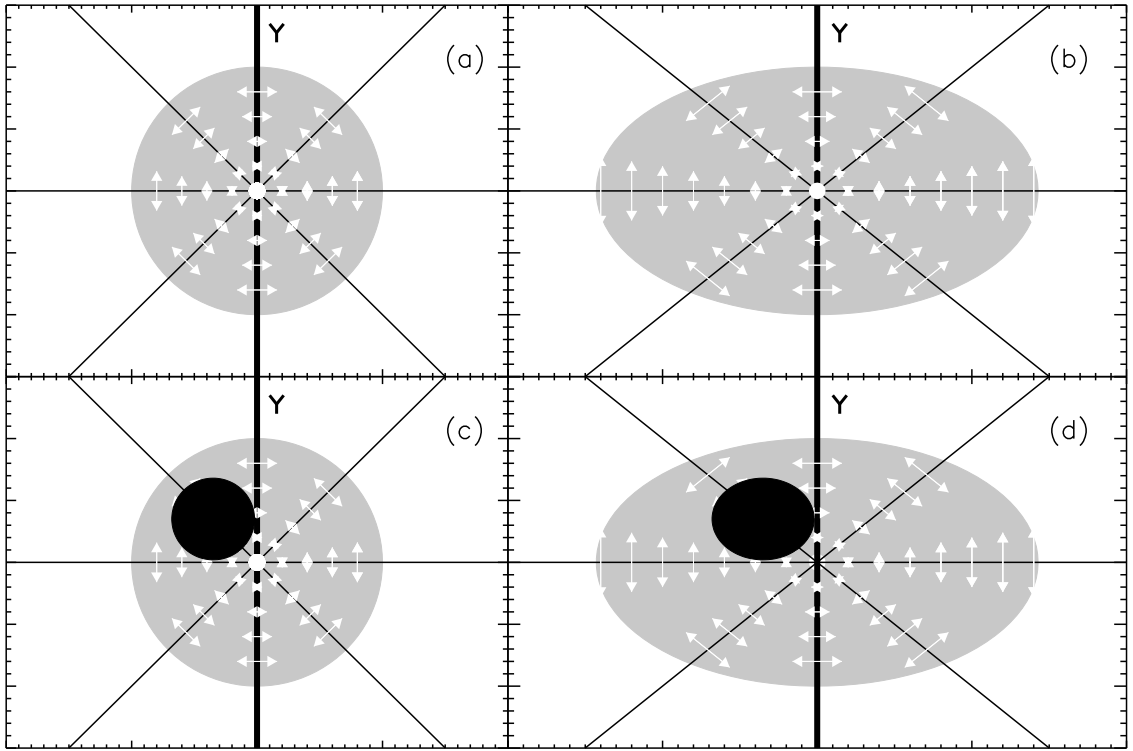}
\caption{The polarization from supernova atmospheres. Each double
arrow in the figure represent a Stokes specific intensity beam
emerging from the photosphere in the observers line of sight. Larger
arrows indicate a higher degree of polarization, not a higher
intensity. The Y-axis is the polarization reference direction.  (a) A
spherical photosphere; the polarization of each beam is exactly
canceled by another one quadrant away so the net polarization is
zero. (b) An ellipsoidal photosphere; vertically polarized light from
the long edge exceeds the horizontally polarized light from the short
edge so $q > 0$. (c) A spherical photosphere with a clump of line
optical depth; the continuum polarization cancels but the
obscuration of diagonally polarized light by the line leads to a
polarization peak feature with $u > 0$ (d) An ellipsoidal photosphere
with a clump of line optical depth; the continuum is polarized in the
q direction and the line in the u direction.
\label{phot_plot}}
\end{center}
\end{figure}

Figure~\ref{phot_plot} illustrates how the polarization of specific
intensity beams emergent from an spherical, pure electron scattering
photosphere might look. The double-arrows indicate the polarization
direction of a beam, with the size of the arrow indicating the degree
of polarization (not the intensity). Note the following two facts: (1)
The polarization is oriented perpendicular to the radial direction.
This follows from nature of the anisotropy of the radiation field. At
all points in the atmosphere (except the center) more radiation is
traveling in the radial direction than perpendicular to it.  Because
the polarization from electron scattering is perpendicular to the
scattering plane, the dominant scattering of radially traveling light
will produce an excess of polarization perpendicular to the radial
direction. (2) The light from the photosphere limb is more
highly polarized than that from the center. This is because the
radiation field at the limb is highly anisotropic -- i.e highly peaked
in the outward (radial) direction.  In addition, photons scattered
into the line of sight from the supernova limb, have generally
scattered at angles closer to 90$^\circ$.

If the projection of the supernova along the line of sight is
circularly symmetric, as in Figure~\ref{phot_plot}a, the polarization
of each emergent specific intensity beam will be exactly canceled by
an orthogonal beam one quadrant away. The integrated light from the
supernova will therefore be unpolarized. A non-zero polarization
measurement demands some degree of asphericity; for example in the
ellipsoidal photosphere of Figure~\ref{phot_plot}b, vertically
polarized light from the long edge of the photosphere dominates the
horizontally polarized light from the short edge. The integrated
specific intensity of Figure~\ref{phot_plot}b is then partially
polarized with $q > 0$. Because an axisymmetric system has only one
preferred direction, symmetry demands that the polarization angle is
aligned either parallel or perpendicular to the axis of symmetry, thus
$u=0$ for the geometry of Figure~\ref{phot_plot}b.

The effect of line opacity on the polarization spectrum can be
complicated.  In general, light resonantly scattered in a line can
become polarized in much the same way as described above for
electrons.  However because randomizing collisions tend to destroy the
polarization state of an atom during an atomic transition, the light
scattered from lines in supernova atmospheres is often assumed
to be completely unpolarized (e.g. \cite{Hoflich-93J} -- we discuss
this assumption in more detail in \S\ref{source_function}). In
ellipsoidal models, it has been shown that the effect of depolarizing
line opacity is primarily to create a decrease in the level of
polarization in the spectrum \citep{Hoflich-93J}.  Because SN~Ia have
more lines in the blue, the polarization in such models typically
rises from blue to red.

In general, however, the fact that a line is depolarizing does not
mean it necessarily produces a decrease in the degree of polarization
in the spectrum. The actual effect will depend sensitively upon the
geometry of the line opacity and the electron scattering medium. For
example, suppose the electron-scattering regime is spherical, but in
an outer, detached layer there is an asymmetric clump of line optical
depth, as shown in Figure~\ref{phot_plot}c. Because the line obscures
light of a particular polarization, the cancellation of the
polarization of the photospheric specific intensity beams will not be
complete. The line thus produces a \emph{peak} in the polarization
spectrum and a corresponding absorption in the flux spectrum. We call
this effect of generating polarization features the \emph{partial
obscuration line opacity effect} or just \emph{partial
obscuration}. In the case of Figure~\ref{phot_plot}c, the clump
primarily absorbs diagonally polarized light, so we expect the
polarization peak to have a dominant component in the u-direction.

A non-axially symmetric supernova is shown in
Figure~\ref{phot_plot}d. The electron scattering medium is
ellipsoidal, so the continuum spectrum will be polarized in the q
direction. The clump of line opacity, which breaks the axial symmetry,
preferentially obscures diagonally polarized light so the line
absorption feature will be polarized primarily in the u direction. As
we see in the next section, this type of two-axis configuration is a
relevant one for SN~2001el.

\subsection{The Polarization of SN~2001el}
\subsubsection{Polarization of The Photospheric Spectrum}
\begin{figure}
\leavevmode
\psfig{file=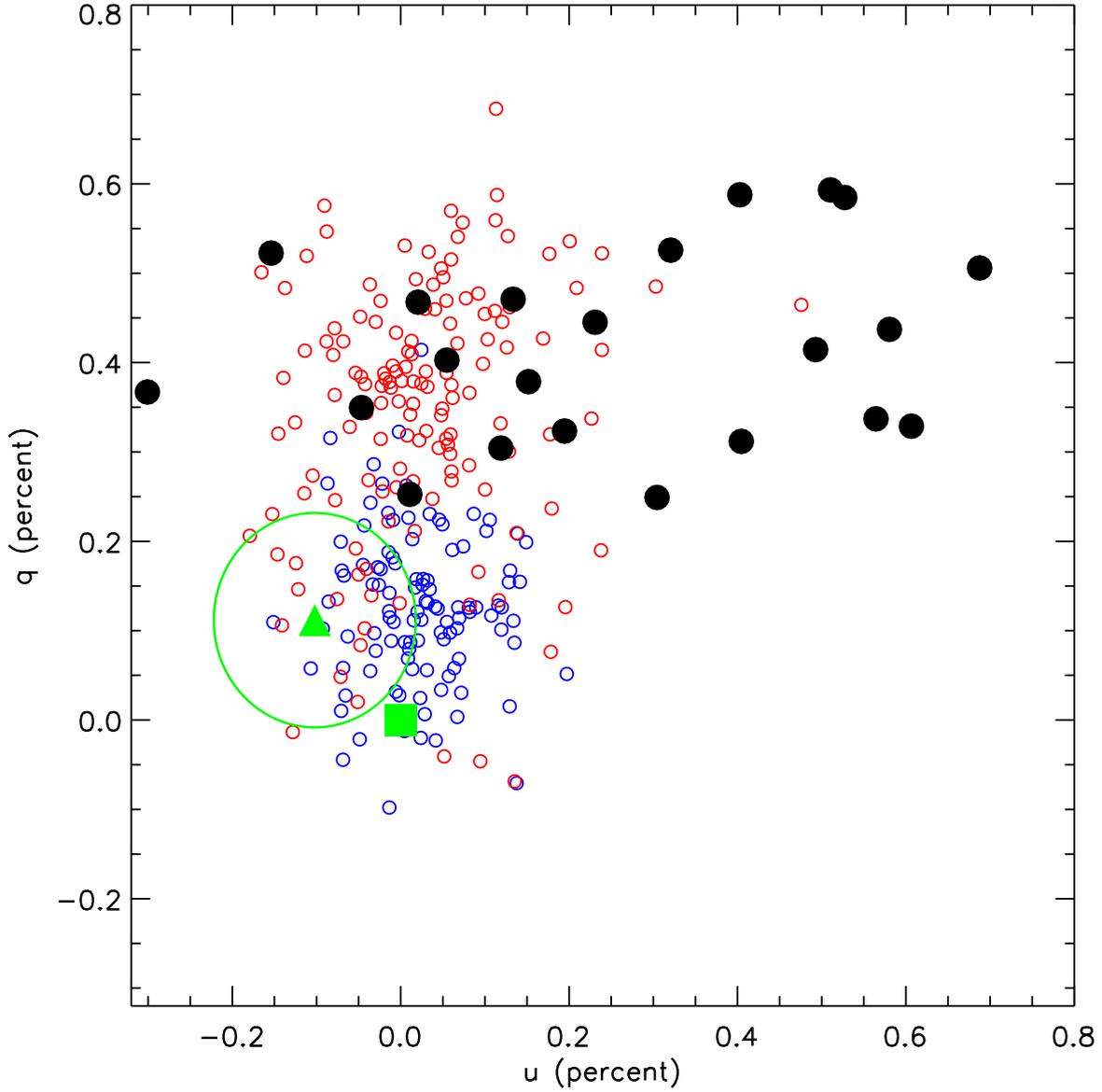,width=6.5in}
\caption{q-u plot of SN~2001el on Sept 25. Each point in the figure
represents a wavelength element of the polarization spectrum. Large
filled circles are points from the HVM feature (7800-8100~\AA).  Small
open circles are points from photospheric spectrum, where the blue
open circles come from the wavelength range (4000-6000~\AA) and the
red ones from (6000-8500~\AA).  The green square at the origin
represents the choice of the ISP leading to the simplest theoretical
interpretation, and the one used in the paper.  The green triangle is
the ISP determined using later time observations and assuming the
intrinsic supernova polarization is zero at this time.  The green
circle is the rough estimated error on the ISP determined in this way.
\label{qu_plot}}
\end{figure}

The q-u plot of SN~2001el is shown in Figure~\ref{qu_plot}.  In order
to interpret the intrinsic supernova polarization, one must first
subtract off the interstellar polarization (ISP), caused by the
scattering of the radiation off aspherical dust grain along the way to
the observer.  The ISP has a very weak wavelength dependence,
\citep{Serkowski} and therefore choosing the magnitude and direction
of the ISP is basically equivalent to choosing the zero point of the
intrinsic supernova polarization in the q-u plane of
Figure~\ref{qu_plot}.  The particular choice of ISP can dramatically
affect the theoretical interpretation of the polarization data (see
\citet{Leonard_98s,Howell-99by}).

The choice of the ISP that leads to the simplest theoretical
description is shown as the green square in Figure~\ref{qu_plot}.  In
this case the photospheric part of the spectrum (open circles), apart
from some scatter, draws out a straight line in the q-u plane --
i.e. the degree of polarization changes across the photospheric
spectrum but the polarization angle remains fairly constant.  This
would be the case if all of the photospheric material followed the
same axial symmetry. The intrinsic polarization spectrum (i.e. percent
polarization versus wavelength) of SN~2001el using this choice of ISP
is shown in Figure~\ref{spec_plot}b.  The degree of polarization rises
from blue to red, as expected in ellipsoidal models due to the higher
line opacity in the blue.  The level of continuum polarization in the
red is about 0.4\%, and the SiII 6150 line represents a depolarization
by about the same amount.  Models of ellipsoidal electron scattering
atmospheres indicate that level of polarization may roughly correspond
to an deviation from spherical symmetry of about $10\%$
\citep{Hoflich-91-pol}.

Although the square in Figure~\ref{qu_plot} is favored by simplicity
arguments, it is preferable to make a direct measurement of the ISP,
if possible.  At late epochs it is believed that the supernova ejecta
becomes optically thin to electron scattering.  The intrinsic
supernova continuum polarization would then be zero, and the observed
polarization due only to the ISP.  Paper~I estimated the ISP in this
way, using observations taken on Nov 9.  Assuming the intrinsic
supernova polarization is zero at this time, the determined ISP (with
an estimated error contour) is shown as the green triangle in
Figure~\ref{qu_plot} .  Although the ISP thus determined is not
grossly inconsistent with the simplest choice, it seems to indicate
that the polarization zero point lies off of the main q-u line. If this is
true, the angle across the photospheric spectrum is no longer
constant.  The photospheric material approximates an axial symmetry,
but an off-axis, sub-dominant component (e.g. a photospheric clump)
must exist to account for the offset of the q-u line.

Because the main purpose of this paper is to explore the geometry of
the HVM, not the photosphere, we will simplify our discussion by
ignoring any off-axis photospheric components.  We will assume the
polarization zero point of the axially-symmetric component is given by
the square and that the photosphere can be approximately modeled as an
ellipsoid.  Although the paricular ISP choice has important
implications for the geometry of the photospheric material, it does
not greatly affect our analysis of the HVM feature.

\subsubsection{Polarization of The HVM Feature}
The HVM flux absorption feature is associated with a polarization peak
in the spectrum (Figure~\ref{spec_plot}b).  Unlike the flux absorption
profile, the polarization peak does not show a clear double feature.
Although the noise of the polarization spectrum makes it difficult to
analyze the line profile, it appears that a peak due to the red
triplet line ($\lambda 8662$) is absent or suppressed compared to the
blue lines ($\lambda 8498$ \& $\lambda 8542$).

In Figure~\ref{qu_plot}, the wavelengths corresponding
to the HVM feature are shown with closed circles. The HVM polarization
angle deviates from the photospheric one, pointing instead mostly in
the u-direction.  The HVM feature also shows an interesting looping
structure -- as the wavelength is increased, the polarization moves
counter-clockwise in the q-u plane.  ``q-u loops'' such as these have
been observed before, for example in the H-alpha feature of SN~1987A
\citep{Cropper-87A}.

The different polarization angle of the HVM feature means that the
geometry of SN~2001el cannot be completely axially symmetric.  The Stokes
$U$ parameter changes sign upon reflecting the system about the
polarization reference axis (see Equation~\ref{stokes_eq}) and
therefore must be zero for any system with a reflective symmetry, such
as the axially-symmetric system of Figure~\ref{phot_plot}b.  The
non-zero u-polarization can not solely be a kinematic effect either,
for although the SN ejecta is expanding, the velocity law is supposed
to be a spherical, homologous one ($v \propto r$) which preserves the
reflective symmetry. As the supernova expands and evolves the density
contours of the system may change as outer layers thin out and reveal
different parts of the underlying material; however unless the
velocity law deviates from homology and shows some preferential
direction, the reflective symmetry will always be preserved and we
must have $u=0$ at all times.  In order to get a non-zero u component,
we must break the reflective symmetry of the geometry with an off-axis
component, such as the clump of Figure~\ref{phot_plot}d.

A natural explanation of the relatively large degree of polarization
and change of polarization angle of the HVM feature is partial
obscuration of polarized photospheric light, somewhat like
Figure~\ref{phot_plot}d. We find in \S\ref{geohvm} that this
interpretation can also account for the q-u loop.  In the next section
we describe a technique for calculating partial obscuration that
allows us to directly compare synthetic polarization spectra to the
data. Other mechanisms could presumably be invoked to explain the HVM
polarization peak, but in this paper we only consider the effects of
partial obscuration.

\section{The Two-Component Polarization Model}
\label{twocomp}
To compute polarization in multi-dimensions most investigators have
employed Monte Carlo methods
\citep{Code-blobs,wood-escatter-I,Hoflich-91-pol}. This approach has
the benefits of generality and ease of coding, but with the drawback
of extreme computational expense. A very large number of photons
must be followed to escape along each line of sight in order to
overcome the random Poisson noise.  This noise must be kept much less
than a fraction of a percent in order to confront the small observed
polarization levels. It is therefore cumbersome to use Monte Carlo
codes in a parameterized way to explore the huge parameter space
available with 3-D geometries.
 
In the case of the HVM, a simplification is possible that allows for a
much faster and more insightful computation. Assuming that the
electron densities in the HVM regime are around $10^{7}
\mathrm{cm}^{-3}$, the optical depth to electron scattering through
the HVM shell is $\tau_{es} = n_e \sigma_t R_{sh} \approx
10^{-3}$. Therefore one can ignore electron scattering in the HVM and
the radiative transfer problem separates naturally into the two
regimes of photosphere and HVM. The photosphere acts as a source of
polarized light illuminating a region of basically pure line optical
depth in the HVM.  Assuming the lines are depolarizing, the only
effect of the HVM is to obscure some of the polarized photospheric
light and re-emit some unpolarized light into the observer's line of
sight.

Because the model makes a sharp distinction between an inner polarized
source (the photosphere) and an outer line-forming region (the HVM),
we call this approach the \emph{two-component model}. The model is
basically a way to formalize the simple pictures of
Figure~\ref{phot_plot}.  The two-component model is constructed to
apply to the detached layers of the HVM. For line forming material
near the photosphere a sharp separation of the two regimes would be
artificial since electron scattering is not entirely negligible in the
line forming region. Because the two-component model does not account
for the multiple scattering between lines and electrons, photospheric
spectra synthesized with it may be incorrect. On the other hand
because the model captures some of the essential features of various
geometries, some qualitative insight may still be gained with respect
to the lines formed near the photosphere. As we are only concerned
with the HVM in this paper, this is not relevant for the present work.

\subsection{The Sobolev Approximation}
The Sobolev approximation is a method for computing line formation in
atmospheres with large velocity gradients. Sobolev models (under the
assumption of a sharp photosphere plus line forming region) have
frequently been used to analyze supernova flux spectra.  Typically
spherical symmetry is assumed (e.g. \citep{Branch_81b, Hatano94D99})
but the method has also been applied in 3D \citep{Rollin_clumps}.
Derivations of the Sobolev method and justification of the
approximation in the modeling of supernova atmospheres can be found in
\citep{Hum_Ryb-78, Castor_1970, Jeff-Branch-Jerusalem}; here we only
quote the important results.

The geometry used in the models is shown in
Figure~\ref{configuration}.  We use a cylindrical coordinate system,
$(p,\phi,z)$ or alternatively a Cartesian one $(x,y,z)$. In either
case the observer line of sight is chosen as the $z$ axis with $z$
\emph{decreasing} toward the observer (i.e. the observer is at
negative infinity). The polarization reference axis is chosen to lie
along the $\phi=0$ (or y) direction, which is also the photosphere
symmetry axis.

For atmospheres in general expansion, such as supernovae, the
wavelength of a propagating photon is constantly redshifting with
respect to the local comoving frame of reference. The insight behind
the Sobolev approximation is that the photon will only interact with a
line in the small region of the atmosphere where the photon is
Doppler-shifted in resonance with the line. The radiative transfer
problem then becomes localized to such ``resonance regions''.  Free
expansion is established in supernova atmospheres shortly after the
explosion; the velocity vector at a point in the atmosphere is in the
radial direction and is given by $\vec{v} = (r - r_0)/t \hat{r}$,
where $r$ is the radius at time $t$ since explosion, and $r_0$ is the
initial radius which is usually small and can be ignored. Consider a
beam of radiation emanating from the photosphere and propagating
through this atmosphere in the $z$ direction, at an impact parameter
$p$ and azimuthal angle $\phi$. Such a beam was illustrated
pictorially as a double-arrow in Figure~\ref{phot_plot}; here we
quantify it with a Stokes specific intensity vector
$\mathbf{I_0}(\lambda,p,\phi)$. If the wavelength of the beam in the
observer frame is $\lambda$, then the wavelength in the local comoving
atmosphere frame is given by the (non-relativistic) Doppler formula:
\begin{equation}
\lambda_{loc} = \lambda \biggl(1 + \frac{\vec{v} \cdot\hat{z}}{c}\biggr)
= \lambda \biggl(1 + \frac{z}{ct}\biggr)
\label{doppler-eq}
\end{equation}
Suppose the only opacity in the atmosphere is due to one line 
with rest wavelength $\lambda_0$. A beam of radiation
will come into resonance with the line when $\lambda_{loc} = \lambda_0$,
which by Equation~\ref{doppler-eq} is at a point:
\begin{equation}
z_r = ct (\lambda_0/\lambda - 1) 
\end{equation}
For each wavelength $\lambda$ in an observed spectrum there is thus a
unique point in the z-direction at which the beam comes in resonance
with the line. According to the Sobolev approximation, the emergent
Stokes specific intensity $\mathbf{I}$ that reaches the observer at
infinity after passing through the line forming region is given by:
\begin{equation}
\mathbf{I}(\lambda,p,\phi) =  \mathbf{I_0}(\lambda,p,\phi)e^{-\tau} +
 (1-e^{\tau}) \mathbf{S}(\lambda,p,\phi,z_r)
\label{SobSpecific}
\end{equation}
where $\tau$ is the Sobolev line optical depth at the point
$(p,\phi,z_r)$ and $\mathbf{S}$ is the Stokes \emph{source-function}
of the line at this point. Both quantities will be explained further
in the sections to come. The first term in Equation~\ref{SobSpecific}
represents photospheric light attenuated by the line optical depth;
the second term represents light scattered or created to emerge into
the line of sight by the line. Equation~\ref{SobSpecific} is identical
to the usual, unpolarized expression for the Sobolev approximation
(see \citet{Hum_Ryb-78}), except now the terms in boldface are all Stokes
vectors.

To generate the observed spectrum of an unresolved object, the
specific intensity of Equation~\ref{SobSpecific} must be integrated
over the projected surface of the atmosphere, i.e. over the $p-\phi$
plane. A wavelength $\lambda$ in the observed spectrum thus gives us
information about the line optical depth and source function
integrated over a plane at $z_r$. Such a plane, which is perpendicular
to the observer's line of sight, is called a constant-velocity (CV)
surface.

In the case of an monotonically expanding atmosphere with more than
one line, a beam of radiation will come into resonance with each line
one at a time, starting with the bluest line and moving to the red. In
this case Equation~\ref{SobSpecific} is readily generalized:
\begin{equation}
\begin{split}
\mathbf{I}(\lambda,p,\phi) = 
 \mathbf{I_0}(\lambda,p,\phi)\exp\biggl(-\sum_{i=1}^{N}\tau_i\biggr)
\\+  \sum_{i=1}^{N}\mathbf{S_i}(\lambda,p,\phi) [1-e^{\tau_i}]
\exp\biggl(-\sum_{j=1}^{i-1}\tau_j\biggr)
\label{SobMultiSpecific}
\end{split}
\end{equation}
where the indices {\it i} and {\it j} run over the lines from red to
blue. Before considering the integration of
Equation~\ref{SobMultiSpecific} over the CV planes, we discuss in more
detail the terms $\mathbf{I_0}$, $\mathbf{S}$, and $\tau$.

\subsection{The Photospheric Intensity}
\label{photosphere}
In this section we calculate the intensity and polarization of specific
intensity beams emergent from an electron scattering photosphere.
We first consider $\mathbf{I_0}(p,\phi)$ in the case that photospheric
regime is spherical (as in Figure~\ref{phot_plot}a) and later show how
to adapt the result to the ellipsoidal case. From the circular
symmetry, the intensity and degree of polarization of a specific
intensity beam can only depend upon the impact parameter $p$ and not
on $\phi$. Let $I_z(p)$ represent the specific intensity in the
$\hat{z}$ direction at $p$, and $P_z(p)$ the degree of polarization of
this beam. The polarized specific intensity is $I_z(p)P_z(p)$ which
will be divided between the $Q$ and $U$ Stokes parameters.

For $\phi = 0$, the polarization points in the horizontal, or negative
$Q$ direction\ -- i.e. $Q(p,\phi=0) = -I_z(p)P_z(p)$ while
$U(p,\phi=0)=0$. The Q and U components at arbitrary $\phi$ are
derived by rotating this expression by $\phi$. The resulting Stokes
vector is:
\begin{equation}
\begin{split}
\bf{I_0} = \begin{pmatrix}I_0 \\ Q_0 \\ U_0 \end{pmatrix}
        = \begin{pmatrix}I_z(p) \\ 
            -P_z(p)I_z(p)\cos(2\phi) \\
            -P_z(p)I_z(p)\sin(2\phi)  \end{pmatrix}
\end{split}
\label{phot_stokes}
\end{equation}
The fact that the trigonometric rotation terms depend on $2\phi$
rather than $\phi$ reflects the fact that the polarization is actually
a quasi-vector \citep{chandra60}.

In the two-component model one must pre-compute the functions $I_z(p)$
and $P_z(p)$. Chandrasekhar first obtained the result for a pure
electron scattering, plane-parallel atmosphere \citep{chandra60}; in
that case $I_z(p)$ follows closely the linear limb darkening law,
while the degree of polarization $P_z(p)$ rises from zero in the
center to $11.2\%$ at the limb; however, the plane-parallel
approximation is not a good one for supernovae, which have extended
atmospheres (i.e. the thickness of the electron scattering zone is a
sizable percentage of its radius). In an extended atmosphere the
radiation field tends toward a more anisotropic distribution, peaking
in the outward direction. This increased anisotropy of the radiation
field leads to generally higher limb polarizations. \citet{Cas-Hummer}
solved the polarized radiative transfer Equation for extended,
spherical electron scattering spheres with density power laws of index
n=2.5 and n=3. They find the polarization can become higher than $50
\%$ at the limb.

We model the photospheric regime as an inner unpolarized boundary
surface, surrounded by a pure electron scattering envelope with a
power law electron density $\rho \propto r^{-n}$. We choose $n=7$, a
density law motivated by SN~Ia explosion models and one that has been
often used in direct spectral analysis \citep{Nomoto_w7,Branch_81b}. The
optical depth (in the radial direction) from the inner boundary
surface to infinity is set at $\tau_{es} = 3$.  The assumption of a
pure electron scattering atmosphere should be a good one for the
wavelength range we are interested in.  The photons that redshift into
resonance with the high velocity IR triplet are those with wavelengths
from 8000-8500~\AA, and there are no strong lines or absorptive
opacities in this region of the spectrum (see \citet{Pinto-Eastman_I}).
At other wavelengths the presence of additional opacities in the
photospheric regime will decrease the polarization from the pure
electron scattering results presented here.

Using a Monte Carlo code, we computed the functions $I_z(p)$ and
$P_z(p)$ for the above scenario. Unpolarized photons were emitted
isotropically from the inner boundary surface. The polarization of
these photons were tracked as they scattered multiple times through
the electron scattering zone. Photons that were back-scattered onto
the inner boundary surface were assumed to be re-absorbed and were
omitted from the calculation. The Monte Carlo code used in this
calculation is a new one developed to further study supernova
polarization in cases where the two-component model is not
applicable. A detailed description of the Monte Carlo code will be
presented in a future paper. We note that the output has been checked
against the results of \citet{chandra60} and \citet{Cas-Hummer}, and
several other cases including \citet{Hillier94} and the analytic
results of \citet{Brown-Mclean}.

\begin{figure}
\begin{center}
\leavevmode
\psfig{file=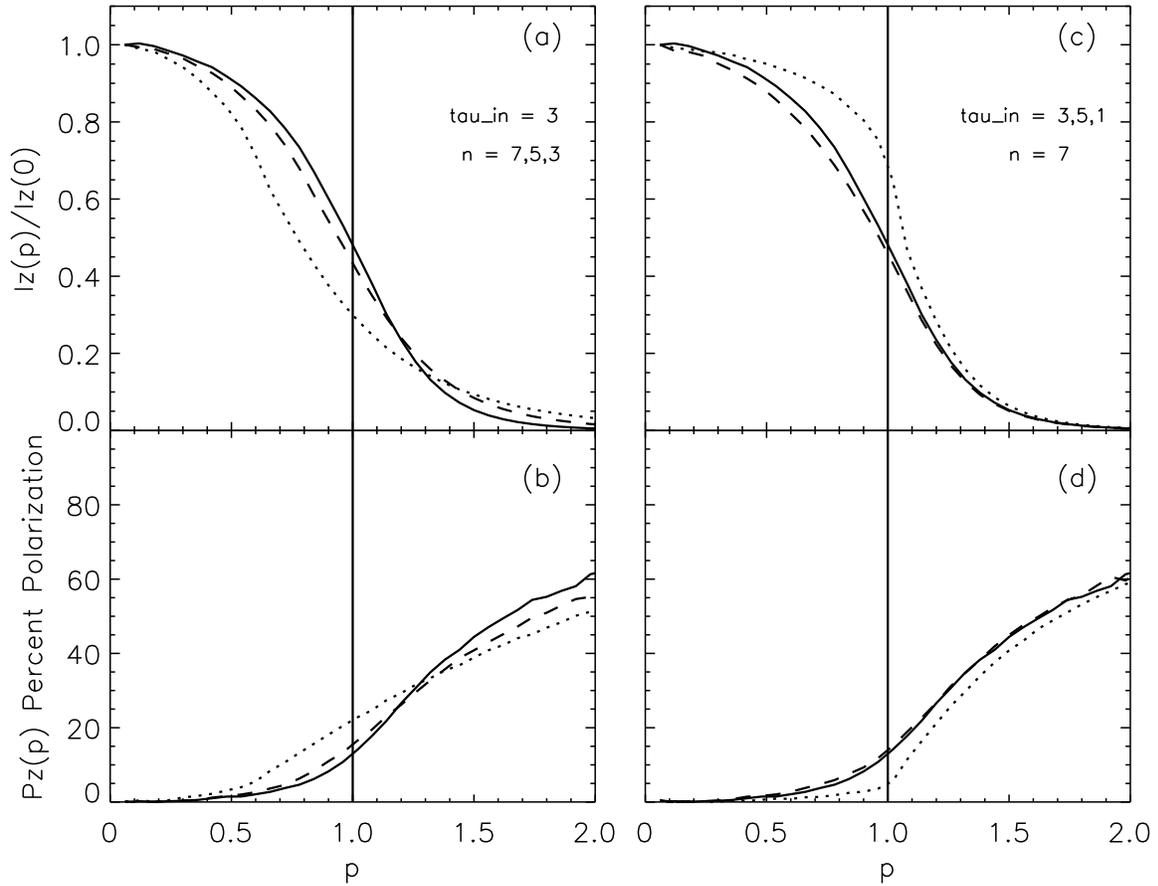}
\caption{The intensity and polarization of specific intensity beams
emerging from the spherical electron scattering photosphere described
in section \S\ref{photosphere}.  The impact parameter p is given in
units of the photospheric radius, defined as continuum optical depth
of one.  The solid lines are the values used in the paper and the
others lines show comparisons with slightly different models. (a,b)
show the dependence on the power law index $n$ assuming $\tau_{ez} =
3$; solid line: n=7, dashed line: n=5, dotted line: n=3. (c,d) show
the dependence of inner optical depth $\tau_{ez}$ assuming $n=7$;
solid line: $\tau_{ez} = 3$,dashed line: $\tau_{ez} = 5$, dotted line:
$\tau_{ez} = 1$.
\label{phot_compare}}
\end{center}
\end{figure}

The computed functions $I_z(p)$ and $P_z(p)$. are shown in
Figure~\ref{phot_compare}. Here $p$ is given in units of the
photosphere radius, defined as the radius at which the optical depth
to electron scattering equals 1.  The intensity and polarization for
$p<1$ do not differ much from the plane-parallel case, with $P_z =
13\%$ at $p=1$.  The photospheric specific intensity does not,
however, terminate sharply at the photospheric radius as is usually
assumed in Sobolev models; rather a significant amount of light is
scattered into the line of sight out to $p \approx 1.4$. Since this
limb light is highly polarized (up to $40\%$) it is important to
include it in our calculations.  Actually most of the polarized flux
comes from an annulus at the edge of the photosphere.  $I_z(p)$ has
become negligible out at the HVM distances of $p \approx 2$, which
confirms that we can make a clear separation between the photospheric
and HVM regimes.

In Figure~\ref{phot_compare} we also compare the $n=7, \tau_{es} = 3$
results to other models with differing density laws and optical
depths. From the similarity of the $n=7$ and $n=5$ models in
Figure~\ref{phot_compare}a and \ref{phot_compare}b it is clear that
the calculations will not depend sensitively on our choice of power
law index. Even if the index were as low as $n=3$, (or worse, not even
described by a strict power law) the behavior of $I_z(p)$ and $P_z(p)$
should still show the same qualitative trends. From
Figure~\ref{phot_compare}c and \ref{phot_compare}d we see the results
also do not depend much on $\tau_{es}$ as long as $\tau_{es} \ga 3$.

The results given so far have not taken into account the asphericity
of the photosphere in SN~2001el.  One could redo the Monte Carlo
calculations for various axisymmetric configurations, but the small
degree of polarization in SN~2001el suggests a rather small ($\sim 10
\%$) deviation from spherical symmetry, so it is not a bad
approximation to apply the spherically symmetric specific intensities
to a slightly distorted photosphere. This technique of using spherical
results to calculate the polarization from distorted atmospheres has
been used, in various manners, by many other authors
\citep{shapsuth,McCall,Jeffery-87A,Cas-Haisch}.

In our models we will only consider the case of an oblate ellipsoidal
atmosphere with axis ratio $E$ and viewed edge-on.  We define an
ellipsoidal coordinate:
\begin{equation}
\eta = \sqrt{x^2 + E^2y^2} 
\end{equation}
Our approximation is that the emergent Stokes intensity from a
position $\eta,\phi$ is given by Equation~\ref{phot_stokes} with
$I_z(p=\eta,\phi=\phi)$ and $P_z(p=\eta,\phi=\phi)$.  In this case we
find an axis ratio of $E \approx 0.9$ is necessary to produce the $0.4\%$
polarization observed in the red continuum of SN~2001el.  This result
agrees with previous, 2-D calculations
\citep{Jeffery-87A,Hoflich-91-pol}.

While the above photospheric model provides a simple and rather
general description of an axially symmetric photosphere, there is no
easy way to assure ourselves that this photospheric model is
unique. The actual specific intensity emergent from an ellipsoidal
atmosphere can depend on the depth and shape of the inner boundary
surface, as well as the inclination of the system.  Moreover, the
polarization of the photospheric spectrum of SN~2001el could arise
from a different kind of asphericity altogether, for instance an
off-center Ni$^{56}$ source, or a clumpy atmosphere. In the absence of
a single preferred photospheric model, we proceed with the above
model, but reiterate that it remains just one of many possible
scenarios. Other choices of $I_z(p,\phi)$ and $P_z(p,\phi)$ must be
investigated on a case by case basis.

\subsection{The Line Optical Depth}
In our synthetic spectra fits, we take the optical depth of the
$\lambda8542$ line, as a free parameter $\tau_1$.  The optical depths
of the other two lines ($\lambda 8662$, $\lambda8498$) are derived
from $\tau_1$.  All three triplet lines come from nearly degenerate
lower levels, so in LTE the relative strength of each line depends
only upon the weighted oscillator strength $gf$ of the atomic
transition.  Even if the level populations deviate from LTE, one
expects the deviation to affect each of the nearly degenerate levels
in the same way. The $\lambda$8542 line has the largest $gf$ value;
$\lambda$8662 is 1.8 times weaker, and $\lambda8498$ 10 times weaker.

\subsection{The Line Source Function}
\label{source_function}
The line source function represents light scattered by the line,
created from the thermal pool or from NLTE effects. Scattering in a
line can polarize light -- as in the case of electron scattering, the
effect is due to the anisotropic redistribution of the different
polarization directions. The angular redistribution depends in general
on the angular momentum {\it J} of the upper and lower levels of the
atomic transition.

Hamilton \citep{Hamilton_47} has considered the linear polarization
from a resonance line, free from collisions. He showed that the
angular redistribution function from such a line could be written as
the sum of an isotropic and dipole term, the relative contributions
depending upon the angular momentum of the transition levels. The
dipole contribution has exactly the same polarizing effect as an
electron scattering, while the isotropic contribution is
unpolarized. The final polarizing effect is thus generally diluted as
compared to the electron scattering case, and can be described by a
polarizabilty factor $W_2$, which varies from 0 for a depolarizing
line to 1 for a line that polarizes like an electron
\citep{Stenflo_94}. Because the Hamilton approach provides a
simple prescription for estimating the intrinsic polarizing effects of
line scattered light, it has often been used outside its scope to
calculate polarized line profiles for non-resonance lines
\citep{Jeffery-87A}.

The Hamilton prescription does not take into account the effect of
collisions.  After a photon has excited the atom, the atom is in a
polarized state with a specific magnetic sublevel {\it M}. If the
collisional timescale is shorter than the lifetime of the transition,
collisions will destroy the polarization state of the atom by
redistributing the atom over all the nearly degenerate magnetic
sublevels, thereby producing an spherically symmetric
configuration. The scattered light will thus be isotropic and
unpolarized. This is the assumption made in the models of
\citet{Howell-99by} (and references therein).

In this paper we use exclusively an isotropic, unpolarized line source
function. In addition to the depolarizing effect of collisions, we
suggest two further reasons why the effect of intrinsic line
polarization is likely a small effect in the case of the HVM feature.
(1) If we evaluate the polarizability factor for the lines of the IR
triplet we find that $W_2$ is almost zero for $\lambda8542$ ($W_2 =
0.02$) and exactly zero for $\lambda8662$.  According to the Hamilton
prescription, only the $\lambda 8498$ line has a moderate polarizing
effect ($W_2 = 0.32$), but this line is by far the weakest of the
three. Note however that since the IR triplet lines are not resonance
lines, the Hamilton prescription does not strictly apply and
complicated NLTE polarizing effects could be operative
\citep{Trujillo-Bueno-99}.  (2) For optically thick lines, photons will
multiple scatter within a resonance region before escaping. On average
the number of scatters in the resonance region is given by $N =
1/P_{esc}$ where the escape probability $P_{esc}$ is given by the
Sobolev formalism:
\begin{equation}
P_{esc} = \frac{1 - e^{-\tau}}{\tau} 
\end{equation}
This multiple scattering has two depolarizing effects: (1) the
radiation field in the line tends toward an isotropic distribution (2)
the probability of the destruction of a photon into the thermal pool
will be increased. For optically thick lines the line-scattered light
will then tend to be unpolarized. On the basis of the spectral fits of
\S\ref{geohvm}, we will argue that the lines of the IR triplet are
saturated ($\tau_1 \ga 5$) for the HVM in front of the photosphere and
thus largely unpolarized.

For an isotropic, unpolarized source function the Stokes vector is:
\begin{equation}
\bf{S} = \begin{pmatrix}S_I \\ S_Q \\ S_U \end{pmatrix}
       = \begin{pmatrix}S_0 \\ 0 \\ 0 \end{pmatrix}
\end{equation}
where $S_0$ is the unpolarized source function.  The actual value of
$S_0$ requires a full NLTE computation of the atomic levels.  For
our purposes a useful parameterization is:
\begin{equation}
S_0 = (1-\epsilon')\bar{J} + \epsilon' B(T)
\end{equation}
The first term represents impinging light scattered by the line, and
so depends upon the mean local radiation field in the line $\bar{J}$;
the second term represents light created from the thermal pool and so
depends upon the Planck function $B$ and the temperature $T$.  The
relative importance of the two factors is governed by $\epsilon'$,
the probability a photon is destroyed into the thermal pool on
traversing the resonance region of a line. In the Sobolev
approximation $\epsilon'$ is given by:
\begin{equation}
\epsilon' = \frac{\epsilon}
{P_{esc} + \epsilon(1 - P_{esc})}
\end{equation}
where $\epsilon$ is the usual static atmosphere destruction
probability.  In NLTE models of supernova atmospheres one finds
$\epsilon$ between 0.05 and 0.1 \citep{nugphd}.  Note as the
probability of a photon's escape ($P_{esc}$) decreases, the chances
that it gets thermalized ($\epsilon'$) increases.

For the value of $\bar{J}$ in the HVM, we use the radiation incident
from the photosphere, ignoring multiple scattering of photons between
the triplet lines (for a discussion of this approximation, see
\citet{Rollin_clumps}).  The photospheric radiation in the HVM is
geometrically diluted by a factor of roughly $\pi r_{ph}^2/ 4\pi
r_{HVM}^2 \approx 1/16$.  Thus for a pure scattering line ($\epsilon'
= 0$), the intensity of the line source function is about 16 times
weaker than the average photospheric intensity.  At the other extreme,
for a thermalized line ($\epsilon' = 1$) and an HVM temperature of
5500 K, the line source function is about 4 times weaker than the
average photospheric intensity.

Because the line source function light is unpolarized and relatively
weak, we find in the end that it has little affect on the synthetic
line profiles.  The exact value of $\epsilon$ is thus not of great
importance.  In our models, we use $\epsilon = 0.01$. 

\subsection{The Integrated Spectrum}
\label{integrated_spectrum}
To obtain the observed Stokes fluxes at a certain wavelength one must
integrate the specific intensity over the CV planes of each line.  For
those CV planes behind the photosphere, we must also account for the
attenuation of the line source function light due to scattering off
electrons as the beam passes through the photospheric region. If we
define $\tau_{es}(p,\phi,z)$ as the electron scattering optical depth
along the z-direction from the point $(p,\phi,z)$ to the observer,
then a fraction $(1 - e^{-\tau_{es}})$ of photons will be scattered
out of the line of sight on their way to the observer.  We assume
these photons are simply removed from the beam and are not
subsequently re-scattered into the line of sight.

For a single line atmosphere, the integrated Stokes fluxes at
wavelength $\lambda$ correspond to material from the CV plane $z_r$
and are given by:
\begin{equation}
\begin{split}
 F_I(\lambda) &= \int\int \biggl[I_z(p,\phi) e^{-\tau}  + \\
  &(1 - e^{-\tau})S_0(p,\phi,z_r)e^{-\tau_{es}}\biggr]pdpd\phi \\
 F_Q(\lambda) &= 
\int\int P_z(p,\phi)I_z(p,\phi)\cos(2\phi) e^{-\tau} pdpd\phi \\
 F_U(\lambda) &= 
\int\int P_z(p,\phi)I_z(p,\phi)\sin(2\phi) e^{-\tau}pdpd\phi
\end{split}
\label{sob_int}
\end{equation}
The integrals can be easily generalized for the case of multiple lines
by applying Equation~\ref{SobMultiSpecific}.

Given our scenario of how the high velocity CaII polarization is
formed by partial obscuration, Equations~\ref{sob_int} give us some
insight into what extent the HVM geometry is constrained by the
polarization measurements. For simplicity, consider the formation of a
single, unblended line, above a spherical photosphere, and suppose we
are trying to reconstruct the distribution of Sobolev line optical
depth $\tau(p,\phi,z)$ over the entire ejecta volume. The Stokes flux
at a certain wavelength gives us information about $\tau$ over the
corresponding CV plane at $z_r$. As Equations~\ref{sob_int}
demonstrate we obviously will not be able to uniquely reconstruct the
distribution of $\tau$ over this plane, because all of the information
gets integrated over to give the three quantities we measure:
$F_I(\lambda),F_Q(\lambda)$, and $F_U(\lambda)$. What we do measure
can be thought of as certain ``moments'' of the $\tau$ distribution
over each CV plane. $F_I$ is a type of ``zeroth moment'', which
depends mostly upon how much material is covering the photosphere,
with little dependence on its geometrical distribution. On the other
hand the $F_Q$ and $F_U$, because of the $\cos2\phi$ and $\sin2\phi$
factors, behave somewhat like ``first moments'', and are sensitive to
how $\tau$ is distributed over the photosphere.  Because the angle
factors $\cos 2 \phi,sin 2\phi$ are rather low-frequency, smaller
scale structures will be averaged out over the integrals, and the
polarization measurements will only constrain the large scale
structures in the HVM.

Before proceeding with the spectral synthesis calculations let us
summarize the assumptions that go into the two-component model. (1)
The electron scattering opacity in the HVM is negligible. (2) the
photospheric regime is reasonably well described by a pure electron
scattering, power law atmosphere, surrounding a finite, unpolarized
source at $\tau_{es} \approx 3$.  (3) For small ($\sim$10 percent)
deviations from sphericity in the photosphere, the angular dependence
of the polarized radiation field does not deviate significantly from the
spherical results (4) The line source function light is unpolarized
(5) Multiple scattering among the triplet lines and between the HVM
and photospheric regime can be ignored.

\section{The Geometry of the High Velocity Material}
\label{geohvm}
\begin{figure}
\begin{center}
\leavevmode
\psfig{file=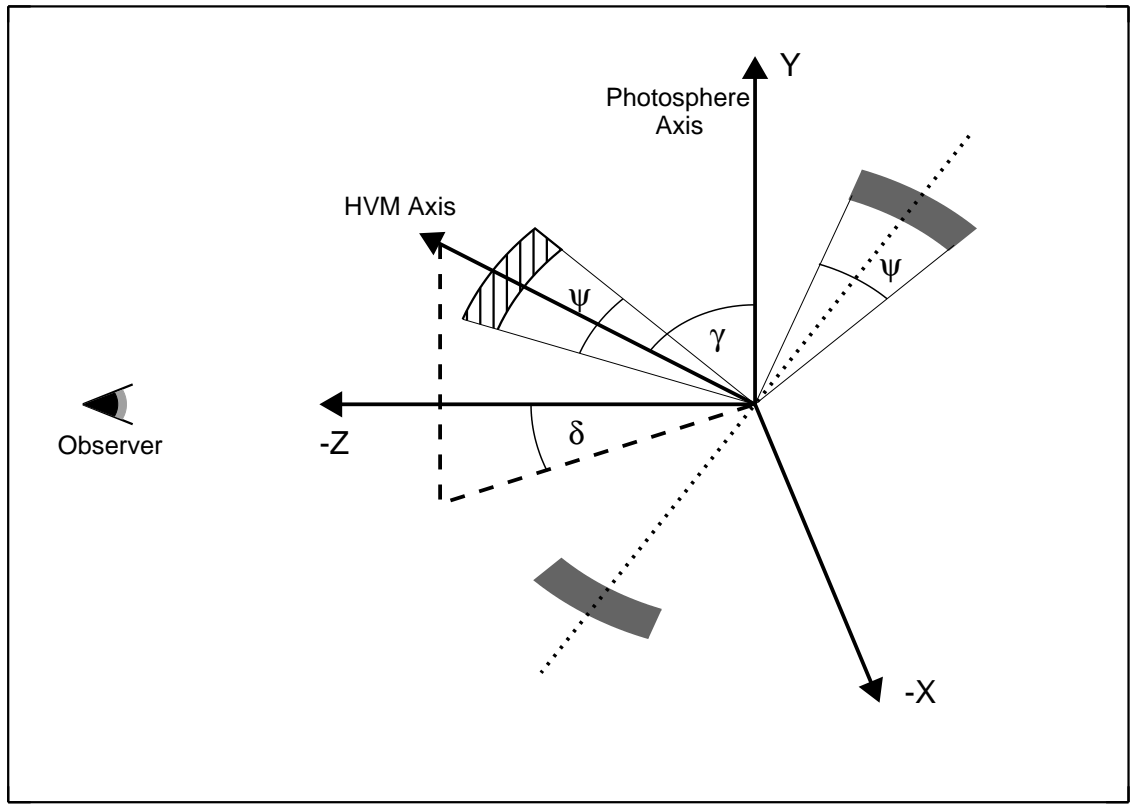}
\caption{ Geometry used in the models.  The line of sight is in the
negative z-direction.  The y-axis is both the polarization reference
direction and the photosphere symmetry axis.  The angles $\gamma$ and
$\delta$ define the orientation of the HVM symmetry axis, where
$\gamma$ is the angle between the y-axis and the HVM axis, and
$\delta$ is the angle between the line of sight and the projection of
the HVM axis onto the z-x plane.  $\psi$ denotes the opening angle of
the clump (hashed arc) and the toroid (solid arc).  The two structures
are generated by spinning the arcs about the HVM axis.
\label{configuration}}
\end{center}
\end{figure}

The speed of the two-component model allows us to explore many
different configurations for the HVM. We report on four possibilities
here, each of which may approximate a structure that is the result of
some particular physical mechanism: (1) A spherically symmetric shell
(2) An ellipsoidal shell with an axis of symmetry rotated from the
photosphere axis of symmetry. (3) A clumped spherical shell (4) A
toroidal structure with a symmetry axis rotated with respect to the
photospheric axis.  The geometry used in the models is shown in
Figure~\ref{configuration}.

The photosphere is modeled as discussed in \S\ref{photosphere}, as an
oblate ellipsoid with axis ratio $E = 0.91$, viewed edge-on.  It is
not the purpose of this paper to explore the detailed geometry of the
photosphere, therefore the ellipsoidal model was chosen as the simplest
possibility that captures the essential features of the
axisymmetric photosphere.  The photosphere symmetry axis is the
y-axis, which is also the polarization reference direction.  The
photospheric intensity is assumed to follow a blackbody distribution
with a temperature $T_{bb} = 9000^\circ K$ chosen to fit the slope of
the red continuum.  We do not attach any physical significance to the
value of $T_{bb}$, but consider it only a convenient fit parameter.

The parameterization of the various HVM geometries is kept simple and
general. The HVM is chosen axially symmetric, with the orientation of
the HVM axis defined by the two angles $\gamma$ and $\delta$.  The
velocities $v_1$ and $v_2$ denote the inner and outer radial
boundaries of the HVM, while $\psi$ is the opening angle (see
Figure~\ref{configuration}).  The reference optical depth $\tau_1$ of
the $\lambda 8542$ line is assumed constant throughout the defined
structure boundaries.  Although this is an idealization of the real
HVM, it allows us to isolate the defining geometrical features of each
structure individually.  Table~\ref{model_table} summarizes the fitted
parameters of each HVM geometry considered in the sections to follow.
Before considering the specific models, we first discuss the general 
constraints that must be met by any HVM model.

\subsection{General Constraints}
\begin{figure}
\begin{center}
\leavevmode
\psfig{file=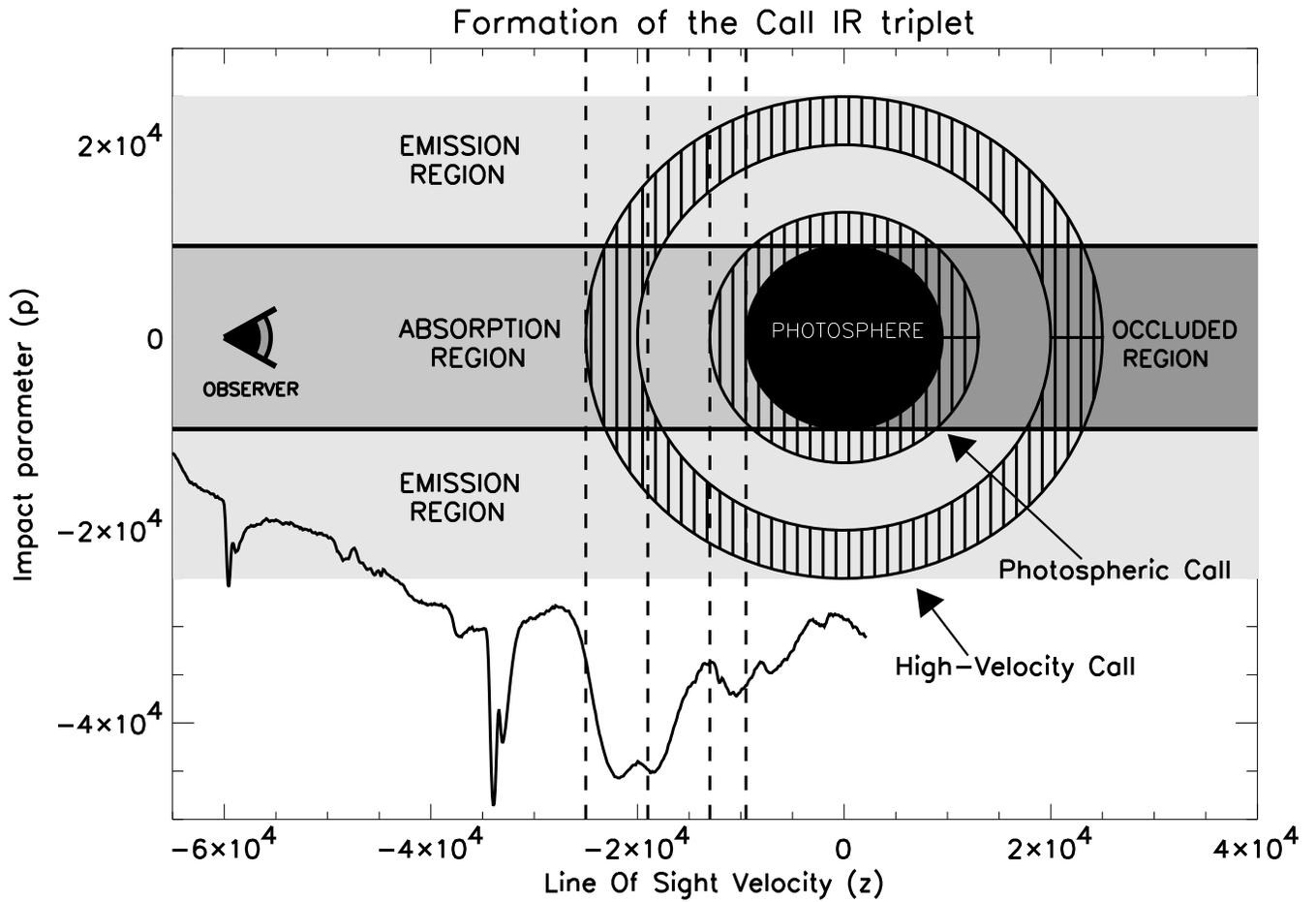}
\caption{Schematic diagram of line formation of the CaII IR triplet
feature in SN~2001el.  The HVM has for illustration been shown with a
spherical shell configuration.  The line profile below is the actual
flux spectrum of the HVM feature on Sept 25.  The vertical lines
represent a few of the CV planes of the $\lambda8542$ line.  Each CV
plane corresponds to unique wavelength in the spectrum, given in the
figure by the wavelength at which they intersect the line profile.
\label{shell_plot}}
\end{center}
\end{figure}

Figure~\ref{shell_plot} is a diagram of the formation of the CaII IR
triplet feature in SN~2001el. The HVM has for illustration been shown
as a spherical shell. The atmosphere can be divided into three
regions, the high-velocity material in each region having a different
affect on the spectrum. (1) The \emph{absorption region}: Material in
the tube directly in front of the photosphere absorbs photospheric
light and emits line source function light into the line of
sight. Since the line source function intensity is usually weaker than
the photospheric intensity, this effect produces an absorption feature
in the spectrum. (2) The \emph{emission region}: material in the outer
lobes does not obscure the photosphere but only adds line source
function light; this produces an emission feature to the red of the
absorption. (2) The \emph{occluded region}: Material in the tube
behind the photosphere is occluded by the photosphere and is not
visible.

Because in our models it is the partial obscuration of polarized
photospheric light that gives rise to the HVM polarization feature,
all of our geometrical information on the HVM will be about the
distribution of CaII in the absorption region. Whether there is any
HVM CaII in the emission region, and if so, what its geometry may be,
will be very difficult to say. In addition we will have absolutely no
information about the material in the occluded region.  In the
spherical HVM shell of Figure~\ref{shell_plot}, about 5\% of the
material is in the absorption region, 5\% is in the occluded region,
and 90\% is in the emission region.  Thus we only probe a small
portion of the potential HVM.  We now consider the general constraints
of these regions in more detail.

\subsubsection{Constraints on the Absorption Region Material}
\label{absorption_region}
We can list 4 general constraints on the HVM absorption region material
that are directly deducible from the Sept. 25 spectra:

(1) The width of the HVM flux absorption feature constrains $\tau_1$
to be non-zero only over the line-of-sight velocity range $18,000-25,000$
\kms. $\tau_1$ is thus confined to a relatively thin region that is
significantly detached from the photosphere.  The edges of the flux
feature are sharp, suggesting that the boundaries of the HVM are
well-defined.

(2) At the minimum of the HVM absorption the flux has decreased by
43\% from the continuum level.  For geometries where the HVM covers
the entire photosphere, the optical depth implied is $\tau \approx
.8$.  On the other hand, some geometries may have higher optical
depths and smaller covering factors, the minimal covering factor being
$f_{min} = 43\%$ for when the lines are completely opaque.  Note that
in this context the term ``covering factor'' denotes the percent of
the photospheric area obscured by the slice of HVM on a plane
perpendicular to the line of sight, corresponding to the resonance
surface of a certain wavelength.  Since this differs from the
traditional usage of the term, we hereafter call this the
\emph{z-plane} covering factor.

We can use the double-dipped flux profile to constrain the z-plane
covering factor of the HVM.  Because the $\lambda$8542 blue triplet
line is intrinsically stronger than the $\lambda$8662 red triplet line
(with a $gf$ value 1.8 times larger), the blue minima of the IR
triplet feature will be about twice as deep the red one \emph{unless}
both lines are saturated.  Because the minima in the HVM feature are
of about equal depth, we conclude that the two lines are indeed saturated
(i.e. $\tau_1 \ga 5$) and the z-plane covering factor is in fact the
minimal one, $f_{min} = 43\%$.

(3) The shape of the flux profile may also constrain the value
$\tau_1$.  Note that two minima in the flux profile have roughly equal
widths.  On the other hand if all three triplet lines are saturated
the blue minima will tend to be wider than the red, due to the
blending of the $\lambda 8498$ with the $\lambda8542$ line.  This
suggests that the $\lambda 8498$ line is weak while the other two
lines are strong, a situation that occurs when $\tau_1 \approx 5$.

(4) Finally, the HVM polarization feature points primarily in the
u-direction. This means the distribution of the HVM is weighted along
the $45^\circ$ line to the photosphere symmetry axis.

\subsubsection{Constraints on the Emission Region Material}
\label{emission_section}
The material in the emission region may be observable as a flux
emission feature to the red of the HVM absorption.  If, for example,
the HVM was a spherical shell, this emission feature would extend from
about $z = -20,000$ to $z = 20,000$, or over 1000~\AA. The emission
from a shell is then very broad, but because the line source function
is much less than the photospheric intensity, the feature is typically
weak and difficult to detect in the spectrum.  A serious problem,
evident from Figure~\ref{shell_plot}, is that the HVM emission feature
overlaps with the photospheric triplet absorption and emission, making
it difficult to separate the two contributions. Only for the HVM
material with $z \ga 15,000$ (i.e. $\lambda > 8700$~\AA) is the HVM
emission feature not blended with the photospheric. Unfortunately the
available spectra of SN~2001el do not extend that far to the red.

The emission region material also affects the polarization level by
diluting the photospheric light with unpolarized line source function
light, thus creating a depolarization feature in the spectrum. Of
course this depolarization feature gives no additional clue as to the
orientation of the emission region material, as the unpolarized line
light carries no directional information. The polarization spectrum of
SN~2001el does have a significant depolarization to the red of the HVM
peak, but since the overlapping photospheric triplet feature may also
depolarize at these wavelengths, it is again not easy to use this to
directly constrain the HVM emission region material.  In our models,
we do not attempt to fit the region redward of 8200 \AA, where the HVM
feature is blended with the photospheric feature.

\begin{figure}
\begin{center}
\psfig{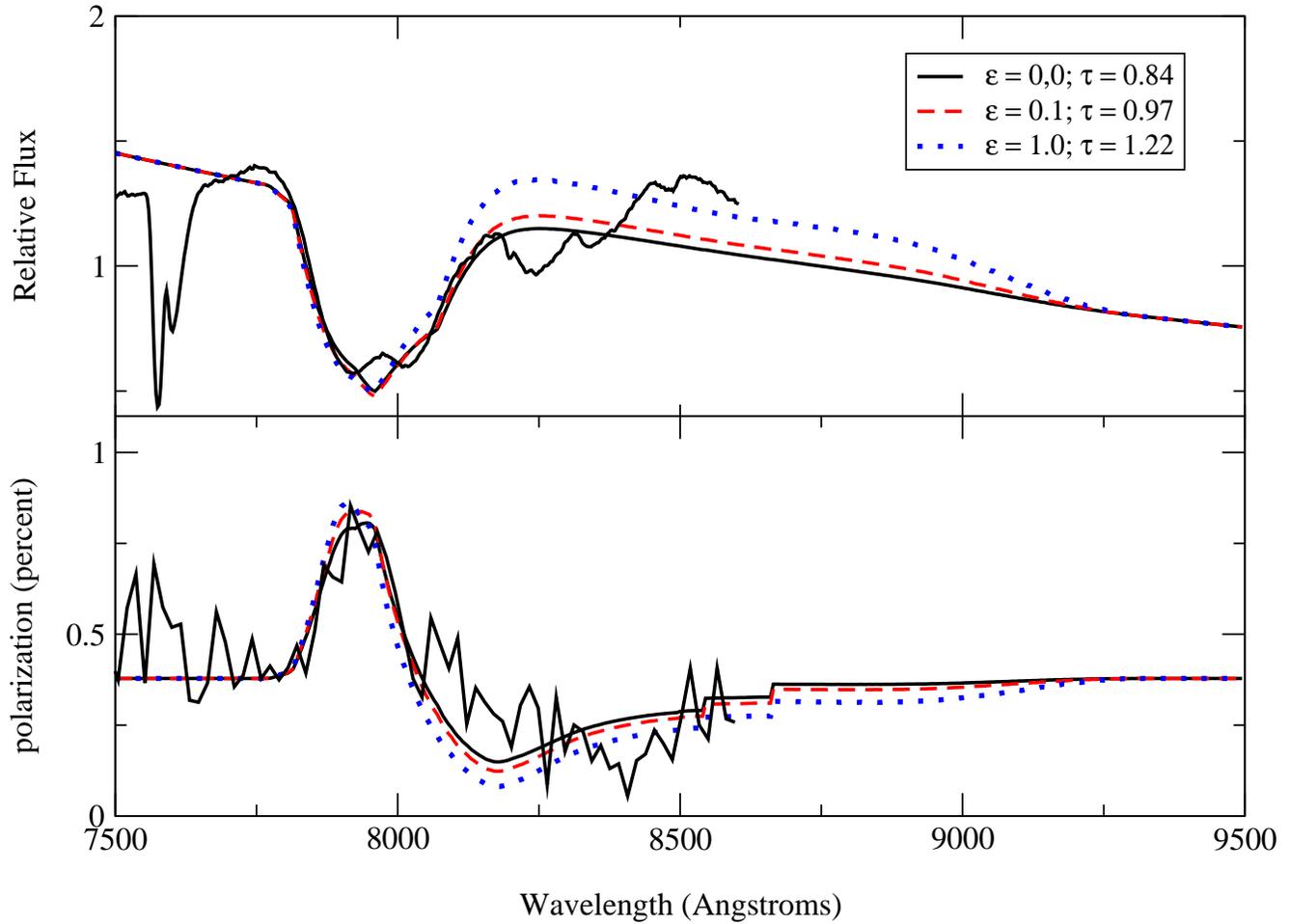}
\caption{The effect of emission region material from a spherical shell
at a temperature $T = 5500^\circ K$.  Note that as the line source
function is increased, the line optical depth must also be increased
in order to reproduce the observed line depth.  A pure scattering line
($\epsilon = 0$; solid line) does not produce a visible emission
feature.  A thermalized line ($\epsilon = 1$; dotted line) produces an
emission, but because this will be blended with the photospheric
triplet absorption and emission, it may still be difficult to detect.
\label{emission_fit}}
\end{center}
\end{figure}

We find that the red emission/depolarization feature is not a very
sensitive diagnostic of emission region material.  The effect on the
spectrum is shown in Figure~\ref{emission_fit} for a spherical shell
with various values of the destruction probability $\epsilon$.  For a
pure scattering line ($\epsilon = 0$) the emission is hardly visible.
For the thermalized line ($\epsilon = 1$) and a temperature $T = 5500
K$, the emission would be substantial but difficult to separate from
the photospheric component.  A value $\epsilon = 1$ is also unlikely
for supernova atmospheres; NLTE models find $\epsilon \approx 0.05$.

The best way to constrain the amount of emission region material is by
line of sight variations (see \S\ref{los}).  The material in the
emission region from one line of sight, becomes material in the
absorption region from another.  With a larger sample of supernovae one
may be able to piece together a picture of the entire volume of high
velocity ejecta.

\subsection{Spherical Shell}
\label{shell_section}
The first HVM geometry we consider is a spherically symmetric shell.
We take the boundaries of the spherical shell to be $v_1 = 20,200$
\kms~ and $v_2 = 25,300$ \kms~ in order to reproduce the line
width. Because the shell curves around, these dimensions actually give
an extension in the z-direction of $18,000-25,000$ \kms, consistent
with constraint (1) of \S\ref{absorption_region}.  The z-plane
covering factor is found to be $\sim 1$, and the optical depth
necessary to fit the line depth $\tau_1 = 0.77$.

\begin{figure}
\begin{center}
\psfig{file=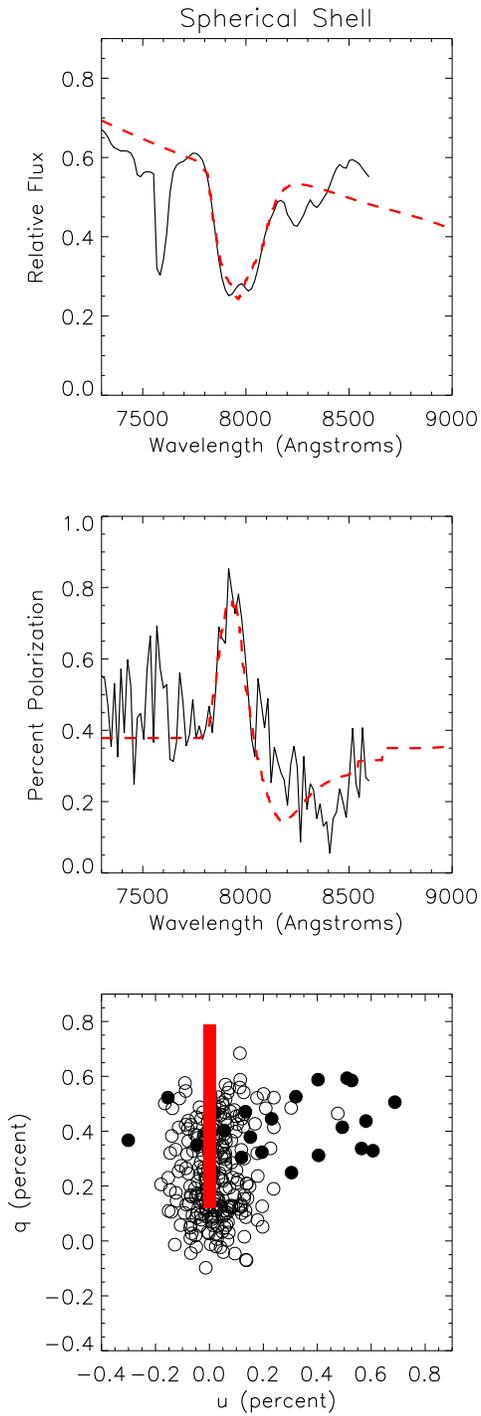,height=7.5in,angle=0}
\caption{Synthetic spectra fits to the observed HVM feature using the
spherical shell model of \S\ref{shell_section}. In the top two plots,
the solid black line is the observed data, and the dashed red line the
fit. In the bottom q-u plot, the black circles are the data and the
red squares the fit. The open circles indicate wavelengths
corresponding to the photospheric spectrum and the solid circles the
HVM feature.
\label{sphere_fit}}
\end{center}
\end{figure}

The triplet lines are not saturated in the spherical shell, so the
model does not satisfy constraint (2) of \S\ref{absorption_region} and
will not well reproduce the flux profile.  In Figure~\ref{sphere_fit}
we compare the synthetic spectra to the observed data. While the
overall fit of the flux feature is decent, the redside minimum is not
well reproduced. We will find better fits to the double-dipped profile
using non-spherical geometries with smaller z-plane covering factors
and saturated lines. Thus the flux spectrum alone suggests a deviation
from spherical symmetry, although the evidence is rather subtle.

\begin{figure}
\begin{center}
\psfig{file=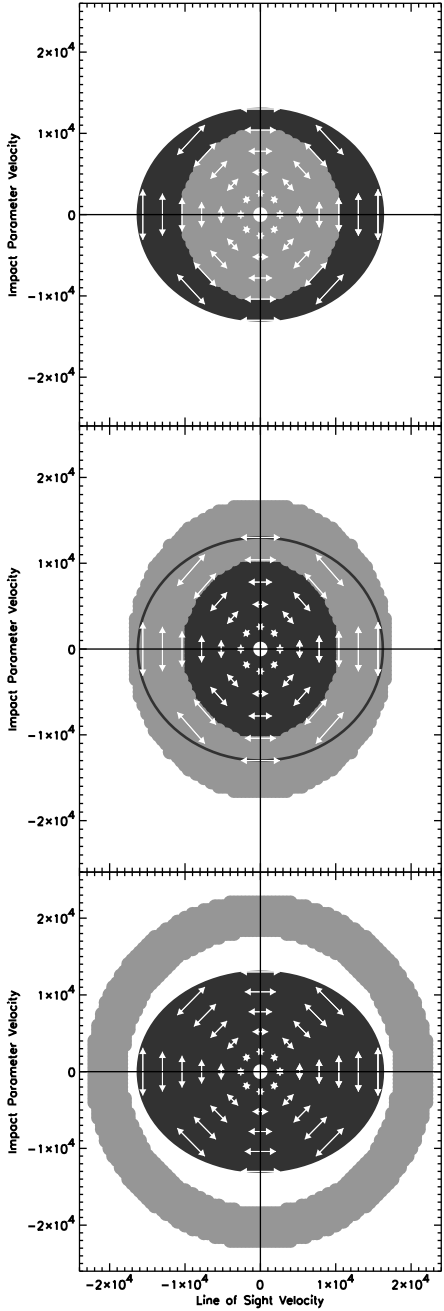,height=7.0in,angle=0}
\caption{Three slices through the spherical shell HVM, which
demonstrate how a detached spherical shell effects the polarization at
three different wavelengths.  Each slice in red is the HVM
cross-section on a plane perpendicular to the z (line of sight) axis,
corresponding to an CV surface for the $\lambda8542$ line at a
particular wavelength. \emph{top}: $v_z = -22,500$~\kms $\rightarrow
\lambda = 7900~\AA$; the line obscures the lowly polarized central
light, leading to a polarization peak \emph{middle}: $v_z =
-15,500$~\kms$\rightarrow \lambda = 8100~\AA$; the line obscures the
highly polarized edge light, leading to a depolarization of the
spectrum \emph{bottom}: $v_z = -5000~$\kms $\rightarrow \lambda =
8400~\AA$; the line does not obscure the photosphere, but since the
line emits some unpolarized line source function light, thus
depolarizing the spectrum.  Note: the photospheric axis-ratio has been
exaggerated (E = 0.8 rather than E=0.91) to clarify the asymmetry.
\label{sphere_slice}}
\end{center}
\end{figure}

The effect of the spherical shell on the polarization is demonstrated
by the slice plots of Figure~\ref{sphere_slice}. At the blue end of
the absorption feature (slice a), the line obscures the weakly
polarized, central light, allowing highly polarized, edge light to
reach the observer. This creates a peak in the polarization
spectrum. Further to the red of the feature (slice b), the line
obscures the edge light and thus depolarizes the spectrum. Even further
to the red (slice c), the line no longer obscures the photosphere, but
the emission region material emits unpolarized line source function
light into the line of sight, and a small level of depolarization
continues. This polarization feature resembles an inverted P-Cygni
profile, as discussed by \citet{Jeffery-Sobolev-P}.

In Figure~\ref{sphere_fit}b we see that the spherical shell naturally
reproduces the correct shape and size of the HVM polarization peak.
The fact that the synthetic polarization feature has only a single
peak is the result of a line blending effect: the red-side
depolarization of the $\lambda$8542 feature suppresses the peak due to
the $\lambda$8662 line.  Note that while the observed depolarization
minimum near 8400 \AA~is not well fit, this is not necessarily a
weakness of the model.  As discussed in \S\ref{emission_section} the
feature at these wavelengths is produced mostly by the calcium near
the photosphere, which has not been included in the model.  In any
case, the spherical shell, which follows the axial symmetry of the
photosphere, does not change the polarization position angle as is
observed (Figure~\ref{sphere_fit}c).  This rules it out as a potential
model for the HVM.

\subsection{Rotated Ellipsoidal Shell}
\label{ellipse_section}
The good fit to the polarization level in Figure~\ref{sphere_fit}
suggests that a shell-like structure may be a viable candidate for the
HVM, as long as the shell is somehow distorted from perfect spherical
symmetry to account for the rotation of the HVM polarization angle.
The simplest scenario is one where the HVM layers of the ejecta are
ellipsoidal with the same oblateness as the photospheric layers, but
with a rotated axis of symmetry. Exactly how such a relative rotation
of the outer layers could arise from an SN~Ia explosion is not
obvious. One might envision that the rapid rotation of a white dwarf
progenitor coupled with a deflagration to detonation transition at an
off-center point \citep{Livne_99} could produce something like this
geometry.

\begin{figure}
\begin{center}
\psfig{file=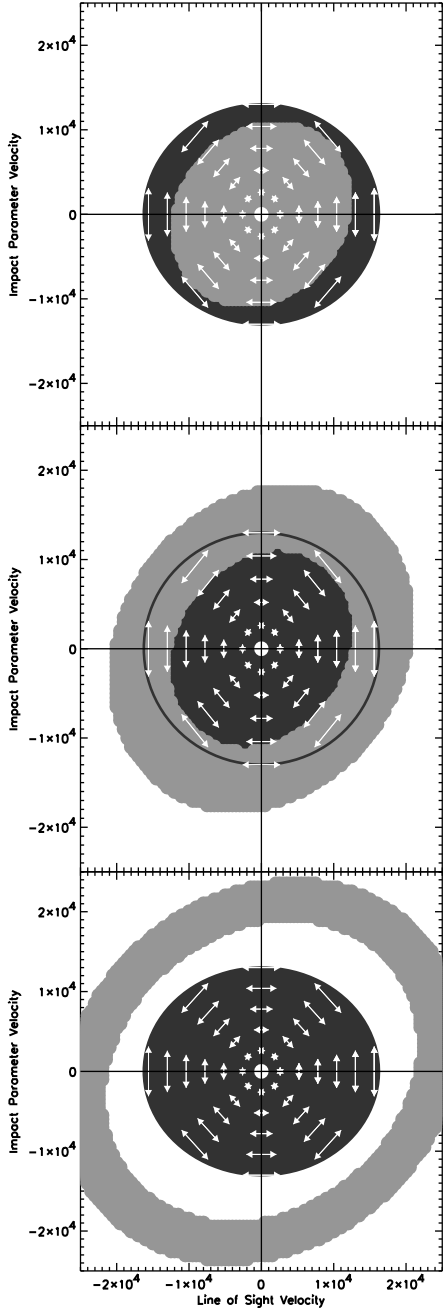,height=7.0in,angle=0}
\caption{Three slices through the rotated ellipsoidal HVM. 
Panels are the same as in Figure~\ref{sphere_slice}.  Because the rotated
ellipsoidal shell preferentially obscures diagonal light, it will produce
a polarization feature with a non-zero u component.  The axis
ratio of both the photosphere and HVM shell are exaggerated (E = 0.8
rather than 0.91) in order to clarify the asymmetries.
\label{ellipse_slice}}
\end{center}
\end{figure}

The effect of the rotated ellipsoidal shell on the polarization
spectrum is demonstrated in the slice plots of
Figure~\ref{ellipse_slice}.  The slices closely resemble those of the
spherical shell (Figure~\ref{sphere_slice}) except that now the
cross-sections of the HVM are ellipses. The shape and size of the flux
and polarization features are thus very similar to the spherical
case. For $\gamma = 0$ (HVM and photosphere axis aligned) the system
is axially symmetric and the HVM polarization feature points in the
q-direction. As $\gamma$ is increased, the ellipses begin to absorb
diagonally polarized light and the HVM polarization feature rotates
into the u-direction.

\begin{figure}
\begin{center}
\psfig{file=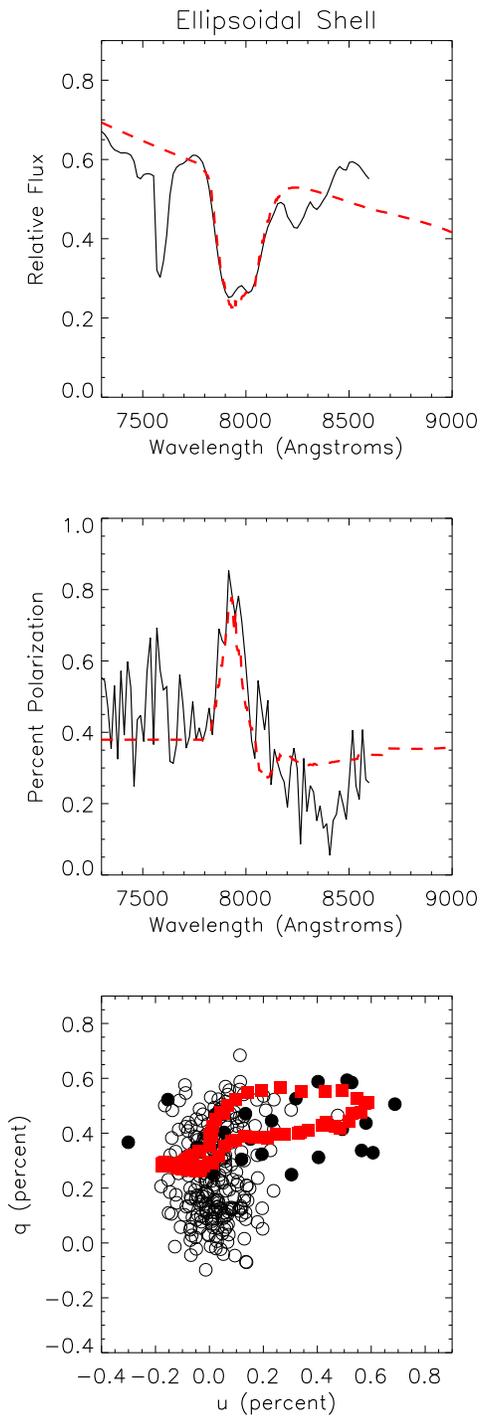,height=7.5in,angle=0}
\caption{Synthetic spectrum fits for the ellipsoidal shell geometry of 
\S\ref{ellipse_section}. The panels are the same as in figure~\ref{sphere_fit}.
The fits to the flux and polarization spectra are similar to the spherical
shell, but now the HVM feature is polarized primarily in the u-direction.
The synthetic feature draws a loop in the q-u plane, which is similar to that
in the observed data.
\label{ellipse_fit}}
\end{center}
\end{figure}

The synthetic spectra for $\gamma = 25^\circ, \tau_1 = 0.77$ are shown
in Figure~\ref{ellipse_fit}.  The ellipsoidal shell, like the
spherical one, fails to meet constraint (2) and does not reproduce the
double-dipped flux profile.  This problem cannot be fixed by changing
the ellipticity of the shell. On the other hand, the ellipsoidal shell
is able to fit the polarization peak and the change of polarization
angle.

Even more interestingly, the ellipsoidal shell produces a q-u loop
similar to that observed in the data.  In our models, we find that a
q-u loop is a common signature of partial obscuration in two-axis
systems. The absorption of the photospheric light typically produces a
peak in both the q and u polarization. The partial obscuration effect
on the q and u polarizations is distinct, so that in general these
features do not peak at the same wavelength, but rather are out of
phase.  When plotted in the q-u plane, this phase offset makes a loop.

\subsection{Clumped Shell}
\label{clump_section}
We parameterize a clumped shell as the section of the spherical shell
lying within a cone of an opening angle $\psi$ (a ``bowl'' shaped
structure, see Figure~\ref{configuration}).  A single clump like this
could perhaps arise if the calcium in the HVM was produced by nuclear
burning that occurred along a preferential axis.  The clumped shell
could also represent one piece of a shell broken into numerous clumps by
an instability, a possibility discussed in more detail at the
end of this section.

In deciding on the appropriate values for the clump parameters, we are
guided by the constraints listed in \S\ref{absorption_region}.  The
opening angle is constrained to $\psi \approx 25^\circ$, so as to
achieve the minimal z-plane covering factor (constraint 2).  The
orientation of the clump axis is chosen so that the clump lies in
between the observer and the photosphere, obscuring the photosphere's
diagonal (constraint 4).

\begin{figure}
\begin{center}
\psfig{file=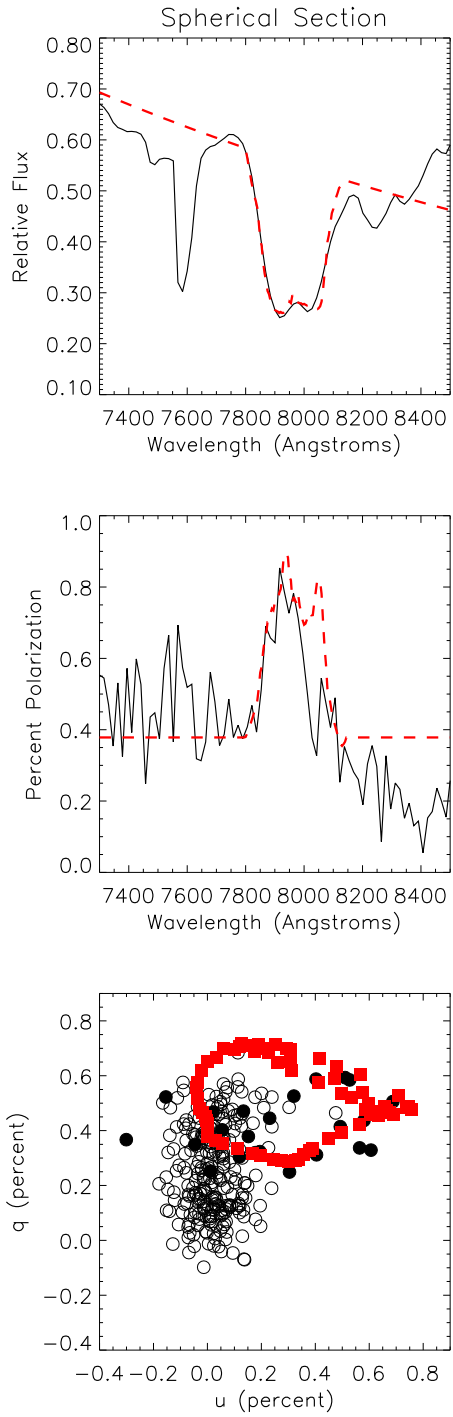,height=7.5in,angle=0}
\caption{Synthetic spectra fits to the HVM feature using the 
clumped shell geometry described in \S\ref{clump_section}.
Panels are the same as in figure~\ref{sphere_fit}
\label{clump_fit}}
\end{center}
\end{figure}

Through trial and error, a reasonable fit to the data was found for
$\psi = 24^\circ, \tau_1 = 5,\gamma = 83.5^\circ, \delta = 4.2^\circ$
The synthetic spectra are shown in Figure~\ref{clump_fit}.  Because
the lines are now saturated, the clump is able to reproduce the two
equal minima of the flux absorption.  The clumped shell also
reproduces the important features of the polarization spectrum --
i.e. the level of polarization, the polarization angle, and the q-u
loop.  On the other hand, the red edges of the synthetic flux and
polarization spectra do not quite match the observed.  In the
polarization spectrum, the peak due to the $\lambda$8662 feature is
not suppressed by blending as it was in the shell models.  This suggests
that our parameterized clump geometry may be too simple and a more
realistic model may involve a complicated superposition of clumps and
shell.

In the geometry described above, the clump axis was chosen almost, but
not quite, perpendicular to the photosphere axis ($\gamma =
83.5^\circ$). One might wonder if the two axes could possibly be
orthogonal ($\gamma = 90^\circ$).  Such a scenario is permissible if
the clump axis remains at an angle $\delta = 4.2^\circ$ to the line of
sight and the whole system is rotated to be observed at an inclination
$i = 90^\circ - 83.5^\circ = 6.5^\circ$.  One might imagine this
geometry as a blob of material that was ejected in the equatorial
direction of the ellipsoidal photosphere.

Although our clumped shell model consists of only a single clump, it
is possible that many more clumps exist in the emission region of the
shell as the extra clumps would leave no obvious signature on the
spectra (see \S\ref{emission_section}).  Clumpiness in a shell could
be caused by various hydrodynamical instabilities.  The expected scale
of such clumpiness is unknown -- it could perhaps take the form of a
single large clump or it could be in the form of numerous smaller
clumps. As we noted in reference to Equation~\ref{sob_int} (see
\S\ref{integrated_spectrum}), the polarization feature due to partial
obscuration is not sensitive to small scale structure, giving rather
the integrated ``moments'' of the optical depth distribution. Thus we
will not be able to empirically constrain the small scale structure of
the clumpiness.  We can say two things though: (1) Whatever the size
of the clumps, their angular distribution must be weighted along the
clump axis defined above. If the clumps were instead small structures
distributed uniformly over the shell, when integrated up they would
average out to the uniform spherical shell analyzed in the previous
section, which did not show a rotation of the polarization angle. (2)
This weighted angular distribution of the clumps cannot vary in the
radial direction. If it did, the polarization angle of the HVM feature
-- which is set by however the randomly placed clumps happen to be
distributed over the photosphere -- would vary randomly across the HVM
feature rather than forming a q-u loop oriented in the
u-direction.  Both of these suggest that the scale of the clumpiness
is not much smaller than the single clump used in the model.

\subsection{Toroid}
\label{toroid_section}
A toroid would be an especially interesting structure to find in the
ejecta of a SN~Ia, as it might give a hint as to the binary nature of
the progenitor system. In the currently preferred progenitor scenarios
(see \citet{Branch_pro_90}), SNe~Ia are the result of a white dwarf
accreting material either from the Roche-lobe overflow of a companion
star or the coalescence with another C-O white dwarf. The orientation
of the accretion disk axis naturally suggests an independent
orientation of the outer ejecta layers, and this could provide a
natural explanation why the HVM of SN~2001el deviates from the
photospheric axis of symmetry.

Whether an accretion disk could maintains a toroidal structure after
the supernova explosion can only be addressed by multi-dimensional
explosion modeling. Here we can calculate what effect such a structure
would have on the flux and polarization spectrum, and whether it could
possibly account for the HVM feature in SN~2001el.  We parameterize
the toroid as the ring of a spherical shell lying within opening angle
$\psi$ (see Figure~\ref{configuration}).

\begin{figure}
\begin{center}
\psfig{file=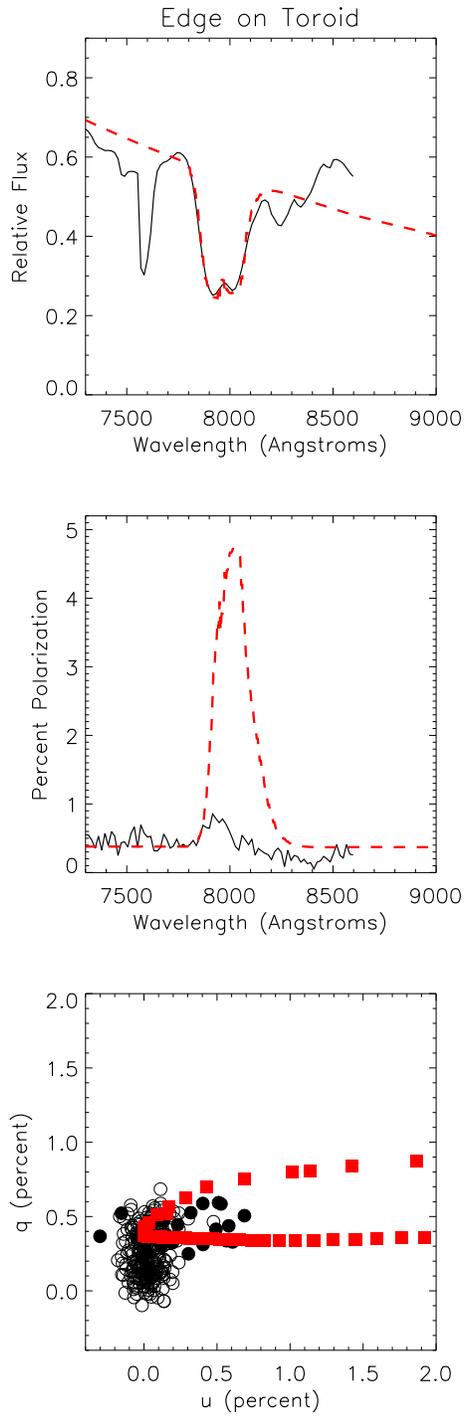,height=7.5in,angle=0}
\caption{Synthetic spectra fits to the HVM feature using the 
edge-on toroid section geometry described in \S\ref{toroid_section}.
Panels are the same as in figure~\ref{sphere_fit}. The polarization feature is
much to strong.
\label{toroid_edge_fit}}
\end{center}
\end{figure}

\begin{figure}
\begin{center}
\psfig{file=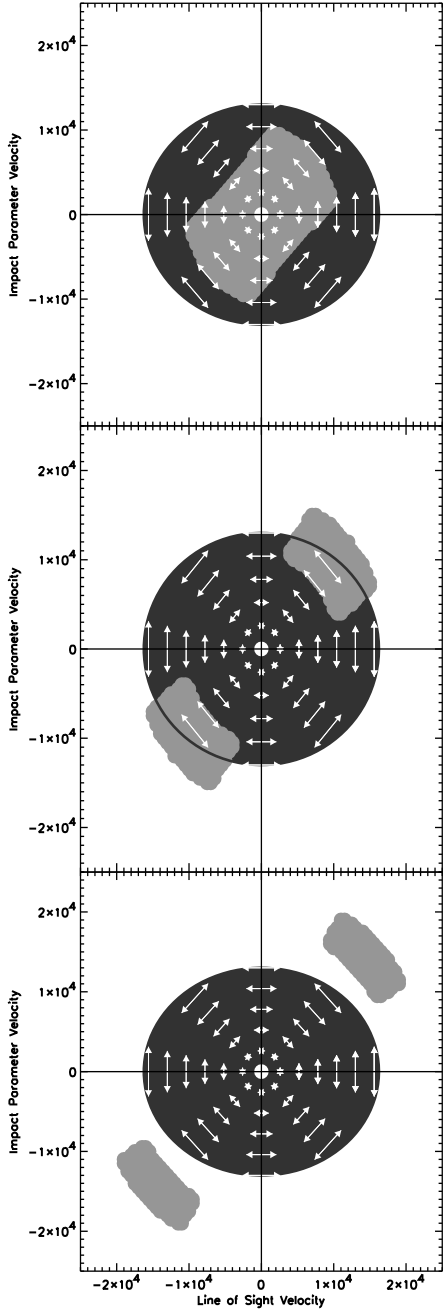,height=7.0in,angle=0}
\caption{Three slices through the edge-on toroid HVM. 
Panels are the same as in Figure~\ref{sphere_slice}.  Because the toroid
is very effective in blocking light of a particular polarization, it
will lead to large polarization peaks.
\label{toroid_slice}}
\end{center}
\end{figure}

We first consider a system where the toroid is observed edge-on.  We
set $\psi = 30^\circ$, giving the minimal z-plane covering factor, and
$\tau_1 = 5$. We orient the torus axis at $\gamma = 45^\circ$ to
preferentially absorb the diagonal light.  The results are shown in
Figure~\ref{toroid_edge_fit}. The flux feature is a good match to the
double-dipped profile, but the polarization peak at 5\% is much too
large. The reason is clear from the slice plot in
Figure~\ref{toroid_slice} -- the edge-on toroid, which occludes
opposite sides of the photosphere, is very effective at blocking light
of a particular polarization.

\begin{figure}
\begin{center}
\psfig{file=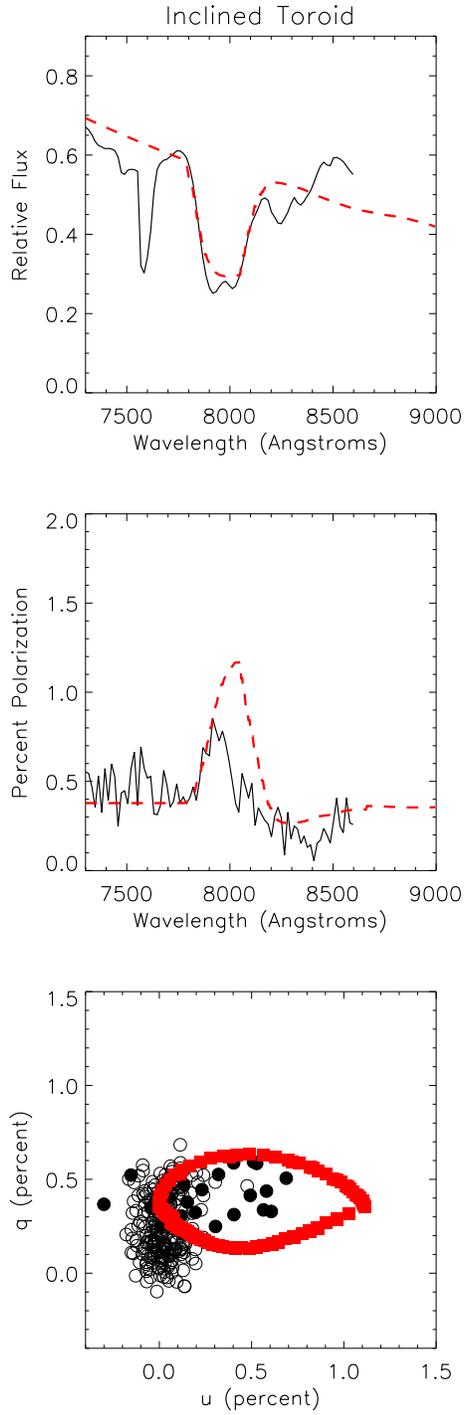,height=7.5in,angle=0}
\caption{Synthetic spectra fits to the HVM feature using the 
inclined toroid geometry described in \S\ref{toroid_section}.
Panels are the same as in figure~\ref{sphere_fit}.  The polarization
feature is still too strong, while the flux absorption is too weak.
\label{toroid_incline_fit}}
\end{center}
\end{figure}

A good fit to the polarization feature can still be sought by changing
the inclination of the toroid. As the inclination is increased, the
toroid rotates off the photodisk and both the flux and polarization
feature decrease. The boundaries of the toroid and the opening angle
must then be readjusted to properly fit the flux feature. In the
present model a perfect fit cannot be found for any inclination. For
all cases where the flux feature is well fit, the polarization feature
is too strong. A compromise fit is shown in
Figure~\ref{toroid_incline_fit}. Here $v_1 = 20,500$, $v_2 = 24,750$
$\gamma = 45^\circ$, $\delta = 43^\circ$, and $\psi = 35 $. The flux
feature is too weak, and the polarization too strong.

\section{The High Velocity Material from Other Lines of Sight}
\label{los}
Previous discussions have pointed out that several different geometrical
configurations are capable of providing reasonable fits to both the
flux and polarization HVM features.  The degeneracy problem is
two-fold: (1) Different distributions of absorbing material in front
of the photosphere can lead to similar polarization features (see the
discussion in \S\ref{integrated_spectrum}) (2) There is no strong
diagnostic of the amount and distribution of material in the emission
region (\S\ref{emission_section}).  In this section we consider how
the degeneracy problem can be overcome by observing the HVM from
multiple lines of sight.

One difficulty in exploring line of sight variations is that the
number of possible configurations in a two-axis system is enormous.
Even holding the boundaries of the HVM fixed, we still have as free
parameters the angle between the photosphere and HVM symmetry axis and
two angles specifying the line of sight. There is no easy way to
catalog all the possibilities. Therefore to keep the discussion simple
and general, in the following calculations we choose the underlying
photosphere to be spherical. The HVM axis can then be aligned in the
z-y plane (i.e. $\delta = 0^\circ$), leaving as the only free
parameter the inclination $\gamma$.  The polarization is then in the q
direction.  Note that in light of Equation~\ref{stokes_eq}, a positive
q-polarization indicates the net flux is vertically polarized, while a
negative q-polarization indicates it is horizontally polarized.

\begin{figure}
\begin{center}
\psfig{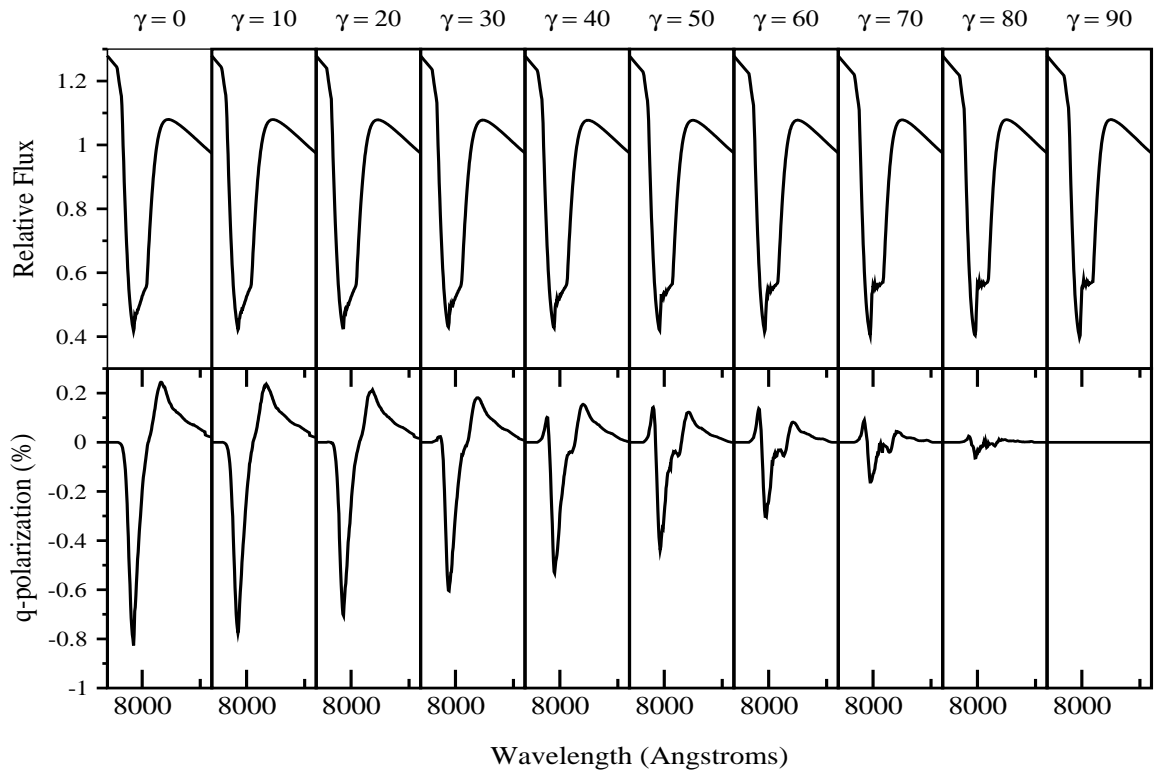}
\caption{Profile from the ellipsoidal shell model along lines of sight
with various inclinations.  Positive (negative) q-polarization
indicates vertically (horizontally) polarized light.  An absorption
feature is visible from all lines of sight.
\label{ellipse_los}}
\end{center}
\end{figure}

The ellipsoidal shell of \S\ref{ellipse_section} shows only subtle
variations with inclination (Figure~\ref{ellipse_los}).  A flux
absorption is visible from all lines of sight, with the absorption
profile barely changing with inclination.  The only effect on the
profile is a small shift of the minimum to the red as the short (i.e
slow) end of the shell moves into the line of sight. For $\gamma =
0^\circ$ (shell viewed edge-on) the polarization is a maximum at
0.8\%; this level is comparable to the HVM feature of SN~2001el.  As
$\gamma$ is increased, the polarization feature decreases
monotonically.  For $\gamma = 90^\circ$ (shell viewed pole on) circular
symmetry is recovered and the polarization is zero.

\begin{figure}
\begin{center}
\psfig{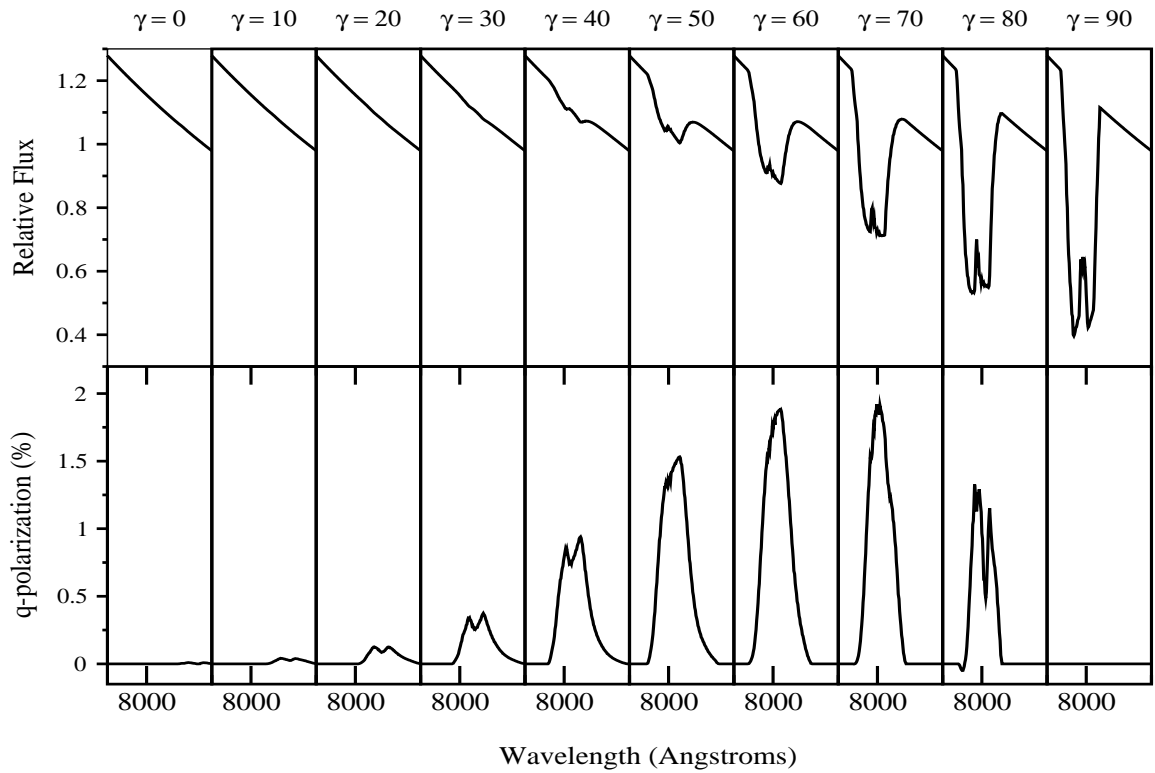}
\caption{Profile from the clumped shell model from various lines of
sight.  As the section moves to the edge of the disk, it blocks lower
intensity, highly polarized edge light.  The flux feature thus gets
weaker while the polarization gets stronger.  Note for $\gamma=40^\circ$
the flux absorption is hardly visible while the polarization feature
is strong.  \label{clump_los}}
\end{center}
\end{figure}

The clumped shell of \S\ref{clump_section}, on the other hand, shows
strong variations with inclination (Figure~\ref{clump_los}). The flux
absorption is deepest for $\gamma=90^\circ$, when the clump is viewed
top on, directly in between the photosphere and observer. At this
inclination, the system is circularly symmetric and the polarization
cancels (the \emph{perfect} cancellation is of course the unnatural
result of our simple ``bowl-like'' clump parameterization; a more
irregularly shaped clump would show a small polarization feature).  As
$\gamma$ is decreased, the clump moves to the edge of the photodisk,
where it covers lower intensity, more highly-polarized light. As a
result, the flux absorption gets weaker while the polarization feature
becomes stronger. A strict inverse relationship holds for the
inclinations $90^\circ-70^\circ$ and provides an important signature
for the single clump model. For inclinations smaller than $60^\circ$
the polarization begins to decrease, but still remains
much stronger than the flux feature. An especially striking signature
occurs for the line of sight $\gamma = 40^\circ$. Here the flux
feature is barely visible while the polarization feature is strong
($\sim 1\%$). The observation of this type of feature would clearly
rule out an ellipsoidal shell and favor a single clump HVM geometry.

The variety of possible flux profiles from the clumped shell model
correspond nicely to the variety of profiles that have already been
observed in some other supernova. As the inclination is decreased from
$90^\circ$, the clump extends further in the z-direction -- the two
lines therefore become broader and the two minima more blended. When
the clump is viewed directly on ($\gamma = 90^\circ$), the two minima
are largely resolved, which is not unlike the feature in SN~2001cx
\citep{li_00cx}. At slightly smaller inclinations ($\gamma \approx
80^\circ$) we found the best fits to the partially blended minima of
SN~2001el. For $\gamma = 40^\circ$ the feature is weaker and the two
minima are almost completely blended, resembling the rounded feature
of SN~1990N \citep{Leib_90n}. For $\gamma = 20^\circ $, the feature is
very weak and about the depth that it was observed in SN~1994D
\citep{meik94d,patat94d}. Thus the clumped shell may be a single model
capable of reproducing the full range of available observations on the
HVM flux feature.  More observations are necessary, however, to
determine if the variety of flux profiles is indeed a line of sight
effect or rather represents individual differences in the high
velocity ejecta.

\begin{figure}
\begin{center}
\psfig{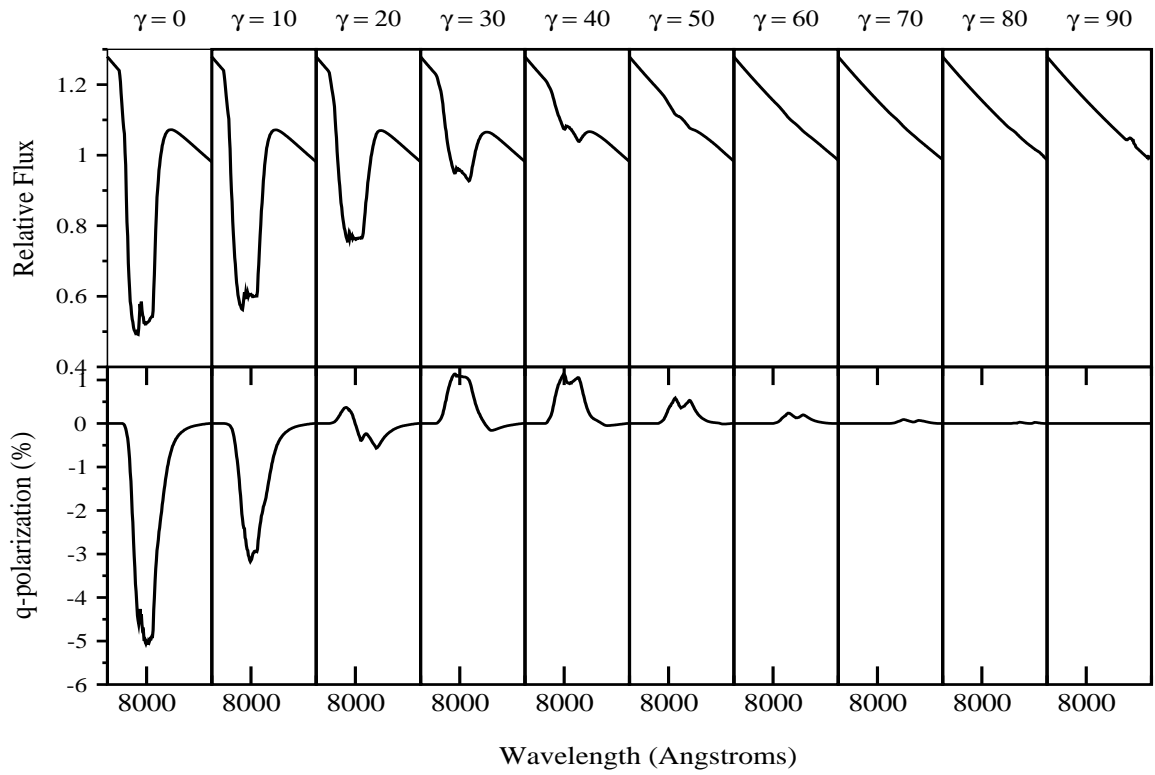}
\caption{Profiles from the toroid model from various lines of sight
\label{toroid_los}}
\end{center}
\end{figure}

The most obvious signature of the toroidal geometry
(Figure~\ref{toroid_los}) is the high levels of polarization ($\sim
5\%$) when viewed near edge-on ($\gamma = 0^\circ$).  An edge-on
toroid occludes vertically polarized light from the edges of the
photosphere, giving a polarization feature with $q < 0$.  As the
toroid is inclined, the structure rotates off the photodisk and both
the flux absorption and polarization peak weaken (in contrast to the
clumped shell model).  At inclinations greater than $20^\circ$, the
toroid begins to occlude the horizontally polarized light from the
bottom of the photosphere -- q then flips sign and becomes positive.

\section{Summary and Conclusions}
High quality spectropolametric observations of supernova may allow us
to extract detailed information on the geometrical structure of the
ejecta.  Interpreting the polarization observations through modeling
is a difficult endeavor, however, largely because of the the enormous
number of configurations available in arbitrary 3-D geometries.  The
huge parameter space and multiple lines of sight make a direct
comparison of data and first principle calculations difficult, not to
mention computationally expensive.  A parameterized approach is
therefore useful in understanding the general polarization signatures
arising from different geometrical structures.  We have taken this
approach here and calculated the polarization features expected from
several geometries potentially relevant to SN~2001el.

The models computed in this paper highlight the wide range of
spectropolametric features possible when aspherical geometries are
considered.  Depolarizing line opacity in the supernova atmosphere
does not in general produce simple depolarization features in the
polarization spectrum.  Asymmetrically distributed line opacity often
creates a polarization peak by partially obscuring the underlying
photosphere.  In systems where the line opacity follows a different
axis of symmetry from the electron scattering medium, the resulting
polarization feature generally creates a loop in the q-u plane.  The
two-component model described in this paper provides a convenient
approach for quickly calculating and gaining intuition into the types
polarization features arising from partial obscuration.

For the case of the high velocity material in type~Ia supernova,
partial obscuration will be a dominant effect on the line features,
resulting in large polarization peaks ($\sim 1$\%) for practically any
geometry considered.  We have therefore explored to what extent
partial obscuration alone can explain the CaII IR triplet polarization
peak in SN~2001el.  Our picture of the SN~2001el ejecta consists of
nearly axially symmetric photospheric material surrounded by a
detached, asymmetric structure at high velocity.  We have investigated
four possible geometries for the HVM: (1) A detached spherical shell
is ruled out because it cannot account for the change of polarization
angle over the HVM feature.  The spherical shell also does not fit the
shape of double-dipped flux absorption profile.  (2) An ellipsoidal
shell, with axis of symmetry rotated $\sim 25^\circ$ from the
photosphere symmetry axis, can account for all the general features of
the HVM polarization spectrum -- the level of polarization, the
polarization angle, and the q-u loop.  However the ellipsoidal shell,
like the spherical one, does not well fit the shape of the flux
absorption profile.  (3) A clumped shell, which could represent a
single clump or a piece of a clumpy shell, can account for all the
general features of the flux and polarization spectra.  (4) A toroid,
in the present model, produces a polarization feature that is larger
than observed.

Different HVM geometries can be clearly discriminated by observing
them from varying lines of sight.  Depending upon the HVM geometry, a
flux absorption similar to that of SN~2001el will be observed in SN~Ia
with different frequency.  For a shell-like model, the flux signature
will be observed from all lines of sight, while for the toroid and
clump, only a fraction of the lines of sight produce the signature
absorption.  Under the assumption that the HVM has a similar structure
in all (or at least a known subset) of SN~Ia's, it may be possible to
constrain the geometry with a statistical sample of early flux
spectra.  Because the different models leave even more dramatic
signatures on the polarization spectra, only a few well-observed
supernova like SN~2001el are needed to discriminate the various
scenarios (see \S\ref{los}).

We have not attempted in this paper to constrain the detailed geometry
of the photospheric material.  Because this material demonstrates a
near axial symmetry, we have adopted the simple and general model of
an edge-on oblate ellipsoid with a power law electron density profile.
The actual photospheric geometry is likely more complicated, and may
deviate from a strict axial symmetry.  Given a more complicated
photospheric structure, one could use the technique described here to
calculate the HVM partial obscuration effect.  Detailed monte-carlo
studies on the structure of the photospheric material are under way;
because the overall asymmetry of the photospheric material is rather
small, however, our main conclusions about the HVM likely hold even
when a more complicated photospheric geometry is used.

Although more observations are necessary to pin down the exact
geometry of the HVM, one can begin to speculate as to its origin.  Two
questions in particular must be addressed: Why is the HVM feature
geometrically detached from the photospheric material?  And: Why does
the HVM deviate from the dominant axis of symmetry of the photospheric
material?

The detachment of the HVM indicates that the atmospheric conditions
change rather suddenly at high velocity.  Three possible changes (or a
combination thereof) could result in an HVM feature (see
\citet{Hatano94D99}): (1) A spike in the overall density in the HVM:
In the SN~Ia deflagration model W7, the material at high velocity
consists of unburnt carbon and oxygen with a solar abundance of
calcium.  The densities of these layers during the epoch in question
are too low to produce an optically thick \ion{Ca}{2} IR triplet.
NLTE models \citep{HVM_mass} show that -- all other things being equal
-- a density increase at high velocity of more than an order of
magnitude is necessary to produce an HVM feature.  (2) A spike in the
calcium abundance: For the W7 densities, the calcium abundance must be
increased by $\sim 10^3$ from solar in order to produce an HVM feature
\citep{HVM_mass}.  This could, for example, be the result of blob of
ejecta material that had undergone explosive oxygen burning, which
increases the calcium abundance by $\sim 10^4$ \citep{Khokolov-93} (3)
A sudden change in the ionization/excitation of the calcium: The
optical depth of the IR triplet is a decreasing function of
temperature (due to the increased ionization of \ion{Ca}{2} to
\ion{Ca}{3}).  Thus it is possible that a temperature decrease in the
outer ejecta layers could make the IR triplet optically thick at high
velocity.  However it seems unlikely in this case that this optical
depth spike would have sharp geometrical boundaries that persisted
over several epochs of observations, as found for SN~2001el.

The distinct orientation of the HVM as compared to the photospheric
material could be (1) the result of random processes in the explosion
physics/hydrodynamics such as Raleigh-Taylor instabilities producing
large scale clumpiness or (2) an indication of a preferred direction
in the progenitor system.  For example, the photospheric dominant axis
could represent the rotation direction of the white dwarf while the
HVM axis could represent the orientation of an accretion disk.
Further explosion and hydrodynamical modeling is necessary to assess
the plausibility of various scenarios.

\acknowledgements PEN and DK acknowledge support from a NASA LTSA
grant.  This research used resources of the National Energy Research
Scientific Computing Center, which is supported by the Office of
Science of the U.S. Department of Energy under Contract No.
DE-AC03-76SF00098. We thank them for a generous allocation of
computing time under the 'Big Splash' award, without which this
research would have been impossible. EB was supported in part by
NASA grants NAG5-3505 and NAG5-12127, and an IBM SUR grant to the
University of Oklahoma.

\clearpage

\begin{deluxetable}{ccccccccc} 
\tablewidth{0pt}
\tablecaption{Fitted parameters for HVM models}
\tablehead{\colhead{Name} & $v_1$\tablenotemark{a}
 & $v_2$\tablenotemark{a}
 & $E$\tablenotemark{b}
 & $\tau_1$\tablenotemark{c}
 & $\psi$\tablenotemark{d} &
 $\gamma$\tablenotemark{e} & $\delta$\tablenotemark{e} 
			& fit-figure}
\startdata
spherical shell   & 20,200 & 25,300 & 1.0  & 0.83  &  - & -   & - &
			\ref{sphere_fit} \\
ellipsoidal shell & 21,200 & 24,800 & 0.91  & 1.20  &  -  &  $25^\circ$ 
			& 90$^\circ$ &	\ref{ellipse_fit} \\
clumped shell     & 20,600 & 24,300 & 1.0  & 5.0   & $23^\circ$ &
         	 	$83.5^\circ$&  $4.2^\circ$ & \ref{clump_fit} \\
edge-on toroid    & 20,900 & 24,500 & 1.0  & 5.0   & $30^\circ$ &
          		$45^\circ$&  $90^\circ$ & \ref{toroid_edge_fit}\\
inclined toroid    & 20,500 & 24,700 & 1.0  & 5.0   & $35^\circ$ &
       			$45^\circ$&  $43^\circ$ & \ref{toroid_incline_fit}\\
\enddata
\tablenotetext{a}
  {$v_1$, $v_2$: inner/outer radial or semi-major boundary in \kms} 
\tablenotetext{b}{$E$: Axis ratio}
\tablenotetext{c}{$\tau_1$: optical depth of reference line ($\lambda8542$)}
\tablenotetext{d}{$\psi$: opening angle (see Figure~\ref{configuration})}
\tablenotetext{e}
  {$\gamma, \delta$: angles defining orientation of HVM symmetry axis 
(see Figure~\ref{configuration})}
\label{model_table}
\end{deluxetable}

\clearpage

\clearpage

\clearpage

\clearpage

\clearpage

\clearpage

\clearpage

\clearpage

\clearpage

\clearpage

\clearpage

\clearpage

\clearpage

\clearpage

\clearpage

\clearpage

\clearpage

\clearpage

\end{document}